\documentclass[11pt]{article}

\usepackage{multicol}
\usepackage{algorithm}
\usepackage{algpseudocode}
\usepackage{enumitem}

\usepackage{graphicx}
\usepackage{times}
\usepackage{url}
\usepackage{multirow}
\usepackage{amsmath}
\usepackage{subfigure}
\usepackage{xcolor}
\usepackage{comment}
\usepackage{url}
\usepackage{booktabs}

\usepackage{cite}
\usepackage{soul}
\usepackage{caption}

\setstcolor{red}

\renewcommand{\paragraph}[1]{\smallskip\noindent {\bf #1}}

\begin{document}

\title{Information Leakage in Encrypted Deduplication via \\ Frequency Analysis:
Attacks and Defenses\footnote{An earlier conference version of this paper appeared at the 47th IEEE/IFIP International Conference on Dependable Systems and Networks (DSN 2017)\cite{li17}. In this extended version, we propose new attack and defense schemes, include a new dataset in our evaluation, and add new prototype experiments.}} 
\author{Jingwei Li$^1$, Patrick P. C. Lee$^2$, Chufeng Tan$^1$, Chuan Qin$^2$, and Xiaosong Zhang$^1$\\
$^1$University of Electronic Science and Technology of China\\
$^2$The Chinese University of Hong Kong \\
Technical Report}
\date{}

\maketitle

\begin{abstract}
Encrypted deduplication combines encryption and deduplication to
simultaneously achieve both data security and storage efficiency.
State-of-the-art encrypted deduplication systems mainly build on deterministic
encryption to preserve deduplication effectiveness.  However, such
deterministic encryption reveals the underlying frequency distribution of the
original plaintext chunks.  This allows an adversary to launch frequency
analysis against the ciphertext chunks and infer the content of the original
plaintext chunks.  In this paper, we study how frequency analysis affects
information leakage in encrypted deduplication, from both attack and defense
perspectives.  Specifically, we target backup workloads, and propose a new
inference attack that exploits {\em chunk locality} to increase the coverage
of inferred chunks.  We further combine the new inference attack with the
knowledge of chunk sizes and show its attack effectiveness against
variable-size chunks.   We conduct trace-driven evaluation on both real-world
and synthetic datasets and show that our proposed attacks infer a significant
fraction of plaintext chunks under backup workloads. To defend against
frequency analysis, we present two defense approaches, namely MinHash
encryption and scrambling.  Our trace-driven evaluation shows that our
combined MinHash encryption and scrambling scheme effectively mitigates the
severity of the inference attacks, while maintaining high storage efficiency
and incurring limited metadata access overhead. 
\end{abstract}

\section{Introduction}
\label{sec:introduction}

To manage massive amounts of data in the wild, modern storage systems employ
{\em deduplication} (see Section~\ref{subsec:dedup}) to eliminate content
duplicates and save storage space.
The common deduplication approach is to store only data copies, called 
{\em chunks}, that have unique content among all already stored chunks.  Field
studies have demonstrated that deduplication achieves significant storage
savings in production, for example, by 50\% in primary storage
\cite{meyer11} and up to 98\% in backup storage \cite{wallace12}.
Deduplication is also adopted by commercial cloud storage services (e.g.,
Dropbox, Google Drive, Bitcasa, etc.) for cost-efficient outsourced data
management \cite{li15}. 

In the security context, combining encryption and deduplication, referred to
as {\em encrypted deduplication} (see Section~\ref{subsec:encrypted-dedup}),
is essential for protecting against information leakage in deduplicated
storage.  Conventional (symmetric) encryption requires that users encrypt data
with their own distinct secret keys.  As a result, duplicate plaintext chunks
will be encrypted into distinct ciphertext chunks, thereby prohibiting
deduplication across different users.  To preserve deduplication
effectiveness, encrypted deduplication requires that each chunk be encrypted
with a secret key derived from the chunk content itself, so that identical
plaintext chunks are always encrypted into identical ciphertext chunks for
deduplication.  Bellare {\em et al.} \cite{bellare13a} propose a cryptographic
primitive called {\em Message-locked encryption (MLE)} to formalize the key
derivation requirement of encrypted deduplication, in which an MLE scheme
consists of a key generation algorithm that maps the content of a message (or
chunk in our case) into a secret key for symmetric encryption/decryption; in
particular, convergent encryption \cite{douceur02} is one classical
instantiation of MLE by deriving the secret key through the hash of a chunk.
On top of MLE, several storage systems address additional security issues,
such as brute-force attacks \cite{bellare13b}, key management failures
\cite{duan14}, side-channel attacks \cite{harnik10}, and access control
\cite{qin17}. 

\begin{figure}[!t]
\centering
\includegraphics[width=3in]{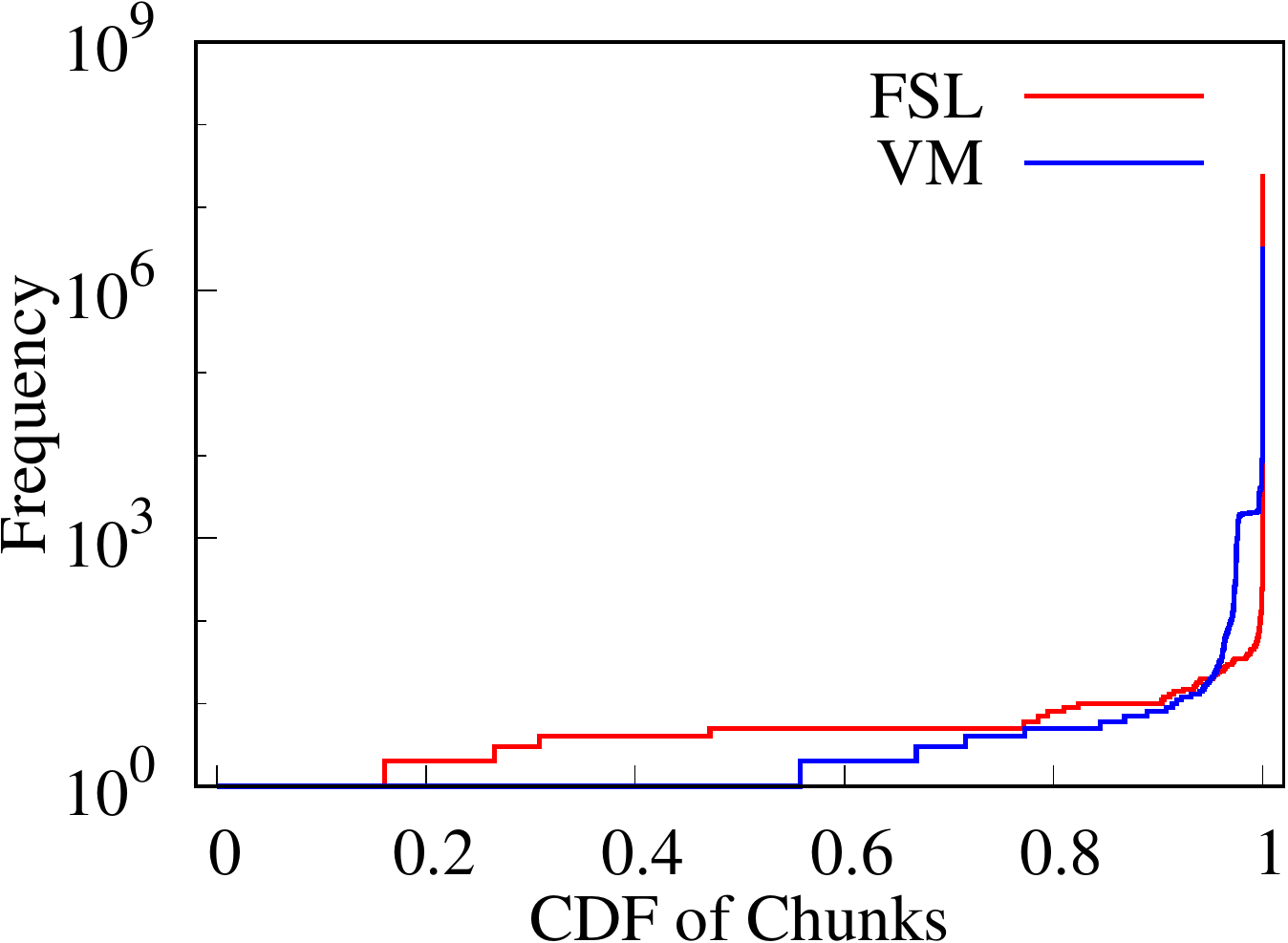}
\caption{Frequency distributions of chunks with duplicate content in the FSL
and VM datasets (see Section~\ref{sec:evaluation} for dataset details).}
\label{fig:freq-dist}
\end{figure}

However, we argue that existing MLE schemes cannot fully protect
against information leakage, mainly because their encryption approaches are 
{\em deterministic}.  That is, each ciphertext chunk is encrypted by a key
that is deterministically derived from the original plaintext chunk.  Thus, an
adversary, which can be malicious users or storage system administrators, can
analyze the frequency distribution of ciphertext chunks and infer the original
plaintext chunks based on {\em frequency analysis}.  We observe that practical
deduplicated storage workloads often exhibit skewed frequency distributions
in terms of the occurrences of chunks with the same content.
Figure~\ref{fig:freq-dist} justifies our observation, by depicting the skewed
frequency distributions of chunks in the real-world FSL and VM datasets (see
Section~\ref{sec:evaluation} for the dataset details).  For example, the FSL
dataset has 99.8\% of chunks occurring fewer than 100 times, while around 30
out of 41~million chunks (or 0.00007\% of chunks) occur over 10,000 times; the
VM dataset has 97\% of chunks occurring fewer than 100 times, while around
15,000 out of 35~million chunks (or 0.04\% of chunks) occur
over 10,000 times.  Such skewed frequency distributions allow the adversary to
accurately differentiate chunks by their frequencies via frequency analysis.
On the other hand, while frequency analysis is a historically well-known
cryptanalysis attack \cite{alkadit92}, the practical implications of frequency
analysis against encrypted deduplication remain unexplored. 

In this paper, we conduct an in-depth study of how frequency analysis
practically affects information leakage in encrypted deduplication.  Our study
spans both attack and defense perspectives, and is specifically driven by the
characteristics of storage workloads in deduplication systems. 

On the attack side, we propose a new inference attack called the 
{\em locality-based} attack, which extends classical frequency analysis to
accurately infer ciphertext-plaintext chunk pairs in encrypted deduplication. 
The main novelty of the locality-based attack is to exploit {\em chunk
locality}, a common property in practical backup workloads.  
Chunk locality states that chunks are likely to re-occur
together with their neighboring chunks across different versions of backups,
mainly because in practice, changes to backups often appear in few clustered
regions of chunks, while the remaining regions of chunks will appear in the
same order in previous backups.  Previous studies have exploited chunk
locality to improve deduplication performance and mitigate indexing overhead
(e.g., \cite{zhu08,lillibridge09,xia11}).  Here, we adapt this idea from a
security perspective into frequency analysis: if a plaintext chunk $M$
corresponds to a ciphertext chunk $C$, then the neighboring plaintext chunks
of $M$ are likely to correspond to the neighboring ciphertext chunks of $C$.  

Our trace-driven evaluation, using both real-world and synthetic datasets,
shows that the locality-based attack can infer significantly more
ciphertext-plaintext pairs than classical frequency analysis.  For example,
for the real-world FSL dataset, the locality-based attack can infer up to
23.2\% of the latest backup data, while the basic attack that directly applies
classical frequency analysis can only infer 0.0001\% of data.  If a limited
fraction (e.g., 0.2\%) of plaintext information of the latest backup is leaked,
the inference rate of the locality-based attack can increase up to 27.5\%.

We further combine the locality-based attack with the knowledge of chunk sizes,
and propose an advanced locality-based attack against variable-size chunks.
The advanced locality-based attack maps ciphertext chunks to some plaintext
chunks with similar sizes, and further increases the inference rate. 

Our inference attacks are harmful in practice, even though the underlying
symmetric encryption remains secure. One security implication of our inference
attacks is that they can identify critical chunks in an encrypted backup
snapshot. Given the plaintext chunks of some critical files (e.g., password
files) in an old backup, an adversary can infer the ciphertext chunks in the
latest backup corresponding to those critical plaintext chunks.  It can then
launch specific attacks against such identified ciphertext chunks; for
example, by dedicatedly corrupting (e.g., deleting or modifying) such
ciphertext chunks, the adversary can make the underlying critical plaintext
information unrecoverable.

On the defense side, we present two defense approaches to combat the
 inference attacks.  The first one is {\em MinHash encryption}, which
derives a common encryption key based on a set of adjacent chunks, such that
some identical plaintext chunks can be encrypted into multiple distinct
ciphertext chunks.  Note that MinHash encryption has been shown to effectively
reduce the overhead of server-aided MLE \cite{qin17}; here we show how it can
also be used to break the deterministic nature of encrypted deduplication and
disturb the frequency ranking of ciphertext chunks.  The second one is 
{\em scrambling}, which randomly shuffles the original chunk ordering during
the deduplication process in order to break chunk locality.  Our trace-driven
evaluation shows that the combined MinHash encryption and scrambling scheme
can suppress the inference rate to only 0.23\% for the FSL dataset.   

We also evaluate the storage efficiency and deduplication performance of the
combined MinHash encryption and scrambling scheme.  First, the combined scheme
maintains the high storage saving achieved by deduplication, and its storage
saving is only up to 3.6\% less than that of the original MLE, which uses
chunk-based deduplication.  In addition, we build a realistic deduplication
prototype based on DDFS \cite{zhu08} and evaluate the on-disk metadata access
overhead.  We show that the combined scheme incurs up to 1.2\% additional
metadata access overhead compared to the original MLE, and it incurs even less
metadata access overhead when there is sufficient memory for metadata caching.
Our findings suggest that the combined scheme adds limited overhead to both
storage efficiency and deduplication performance in practical deployment,
while effectively defending against frequency analysis. 

The source code of our attack and defense implementations as well as the
deduplication prototype is available at: 
{\bf \url{http://adslab.cse.cuhk.edu.hk/software/freqanalysis}}.

The remainder of the paper proceeds as follows. 
Section~\ref{sec:basics} reviews the basics of encrypted deduplication and
frequency analysis.
Section~\ref{sec:threat} defines the threat model. 
Section~\ref{sec:attack} presents our proposed inference attacks based on
frequency analysis.
Section~\ref{sec:evaluation} presents the evaluation results of our proposed
inference attacks.
Section~\ref{sec:defense} presents the defense schemes against the inference
attacks.
Section~\ref{sec:defense_eval} presents the evaluation results of our defense
schemes. 
Section~\ref{sec:related} reviews the related work, and finally,
Section~\ref{sec:conclusion} concludes the paper. 

\section{Basics}
\label{sec:basics}

\subsection{Deduplication}
\label{subsec:dedup}

Deduplication can be viewed as a coarse-grained compression technique to save
storage space.  We focus on chunk-based deduplication that operates at the
granularities of chunks.  Specifically, a deduplication system
partitions input data into variable-size chunks through content-defined
chunking (e.g., Rabin fingerprinting \cite{rabin81}), which identifies chunk
boundaries that match specific content patterns so as to remain robust against
content shifts \cite{eshghi05}.  We can configure the minimum, average, and
maximum chunk sizes in content-defined chunking for different granularities. 
After chunking, each chunk is identified by a {\em fingerprint},
which is computed from the cryptographic hash of the content of the chunk.
Any two chunks are said to be identical if they have the same fingerprint, and
the collision probability that two non-identical chunks have the same
fingerprint is practically negligible \cite{black06}.  Deduplication requires
that only one physical copy of identical chunks is kept in the storage system,
while any identical chunk refers to the physical chunk via a small-size
reference.  

To check if any identical chunk exists, the deduplication system maintains a
\emph{fingerprint index}, a key-value store that holds the mappings of all
fingerprints to the addresses of physical chunks that are currently stored.
For each file, the storage system also stores a \emph{file recipe} that lists
the references to all chunks of the file for future reconstruction. 

\subsection{Encrypted Deduplication}
\label{subsec:encrypted-dedup}

Encrypted deduplication ensures that all physical chunks are encrypted for
confidentiality (i.e., data remains secret from unauthorized users and even
storage system administrators), while the ciphertext chunks that are
originated from identical plaintext chunks can still be deduplicated for
storage savings.  As stated in Section~\ref{sec:introduction}, message-locked
encryption (MLE) \cite{bellare13a} is a formal cryptographic primitive for
encrypted deduplication, in which each chunk is encrypted/decrypted by a
secret key that is derived from the chunk content itself through some key
generation algorithm.  For example, convergent encryption \cite{douceur02} is
one popular MLE instantiation, and uses the cryptographic hash of a chunk as
the corresponding symmetric key.  This ensures that identical plaintext chunks
must be encrypted into the identical ciphertext chunks, thereby preserving
deduplication effectiveness.  Note that the encrypted deduplication system
needs to maintain a {\em key recipe} for each user to track the per-chunk keys
for future decryption.  Each key recipe is encrypted by the user's own secret
key via conventional encryption for protection (see Section~\ref{sec:threat}). 

MLE is inherently vulnerable to the offline brute-force attack
\cite{bellare13a}, which allows an adversary to determine which plaintext
chunk is encrypted into an input ciphertext chunk.  The brute-force attack
works as follows.  Suppose that the adversary knows the set of chunks from
which the underlying plaintext chunk is drawn. Then for each chunk from the
set, the adversary finds the chunk-derived key (whose key derivation algorithm
is supposed to be publicly available), encrypts the chunk with the
chunk-derived key, and finally checks if the output ciphertext chunk is
identical to the input ciphertext chunk.  If so, the plaintext chunk is the
answer.  Thus, MLE can only achieve security for {\em unpredictable} chunks
\cite{bellare13a}, meaning that the size of the set of chunks is sufficiently
large, such that the brute-force attack becomes infeasible. 

To protect against the brute-force attack, DupLESS \cite{bellare13b} 
realizes \emph{server-aided MLE}, which outsources MLE key
management to a dedicated {\em key manager} that is only accessible by
authenticated clients.  Each authenticated client needs to first query the key
manager for the chunk-derived key.  Then the key manager computes and returns
the key via a deterministic key derivation algorithm that takes the inputs of
both the chunk fingerprint and a system-wide secret maintained by the key
manager itself.  This makes the resulting ciphertext chunks appear to be
encrypted by random keys from the adversary's point of view.  In
addition, the key manager limits the rate of key generation to slow down any
online brute-force attack for querying the encryption keys.   If
the key manager is secure against adversaries, server-aided MLE ensures security
even for predictable chunks; otherwise, it still maintains security for
unpredictable chunks as in original MLE \cite{bellare13a}. 

Most existing MLE implementations, either based on convergent encryption or
server-aided MLE, follow {\em deterministic encryption} to ensure that
identical plaintext chunks always form identical ciphertext chunks to
make deduplication possible.  Thus, they are inherently vulnerable to
frequency analysis as shown in this paper.  Some encrypted deduplication
designs are based on non-deterministic encryption
\cite{abadi13,bellare13a,bellare15,liu15}, yet they still keep deterministic
components \cite{bellare13a}, incur high performance overhead \cite{liu15}, or
require cryptographic primitives that are not readily implemented
\cite{abadi13,bellare15}.  We elaborate the details in
Section~\ref{sec:related}. 

\section{Threat Model}
\label{sec:threat}

\subsection{Overview}

We focus on backup workloads, which have substantial content redundancy and
are proven to be effective for deduplication in practice
\cite{zhu08,wallace12}.  Backups are copies of primary data (e.g., application
states, file systems, and virtual disk images) over time.  They are typically
represented as weekly full backups (i.e., complete copies of data) followed by
daily incremental backups (i.e., changes of data since the last full backup),
while the recent trend shows that full backups are now more frequently
performed (e.g., every few days) in production \cite{amvrosiadis15}.  Our
threat model focuses on comparing different versions of full backups from the
same primary data source at different times.  In the following discussion, we
simply refer to ``full backups'' as ``backups''. 

\begin{figure}
\centering
\includegraphics[width=5in]{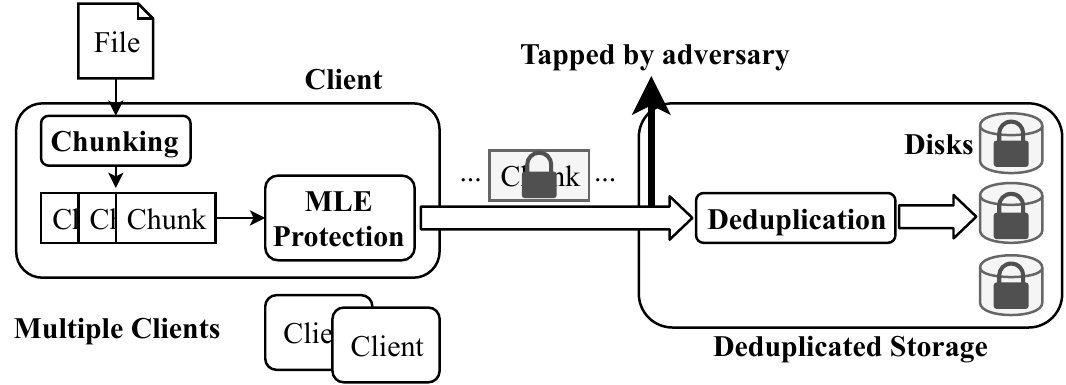}
\caption{Architecture of encrypted deduplication.}
\label{fig:architecture}
\end{figure}

Figure~\ref{fig:architecture} shows the encrypted deduplication architecture
considered in the paper.  Suppose that multiple clients connect to a shared
deduplicated storage system for data backups.  Given an input file, a client
divides file data into plaintext chunks that will be encrypted via MLE.  It
then uploads the ciphertext chunks to deduplicated storage.  An adversary can
eavesdrop the ciphertext chunks before deduplication and launch frequency
analysis.  We assume that the adversary is honest-but-curious, meaning that it
does not change the prescribed storage system protocols or modify any stored
data. 

\subsection{Auxiliary Information}

To launch frequency analysis, the adversary should have access to the
{\em auxiliary information} that provides ground truths about the backups
being stored.  Prior studies have proposed different approaches to obtain the
auxiliary information to launch inference attacks.  We briefly discuss several
representative ones that inspire our work. 
\begin{itemize}[leftmargin=*]
\item 
Naveed {\em et al.} \cite{naveed15} examine inference attacks against the
encrypted databases for electronic medical records, some of which are
protected by deterministic encryption.  To evaluate the feasibility of
launching inference attacks, the authors obtain the auxiliary information from
a public user dataset released by the government health services. 
\item 
Grubbs {\em et al.} \cite{grubbs17} infer the plaintexts of the attributes of
customer records (e.g., first name, last name, ZIP codes, birth dates, etc.)
stored in an encrypted database.  They obtain the auxiliary information
regarding the plaintext distribution via the public census and survey datasets. 
\item 
Bindschaedler {\em et al.} \cite{bindschaedler18} also infer the plaintexts of
the attributes of encrypted customer records like Grubbs {\em et al.}
\cite{grubbs17}, but use the public and purchased U.S. voter registration
lists as the auxiliary information. The authors also use the older versions
of purchased hospital-discharge data and public censor data to infer the newer
versions of respective data.
\item 
Grubbs {\em et al.} \cite{grubbs16} focus on Ubuntu Internet Relay Chat (IRC)
logs \cite{uthus13,irc} and extract the log keywords.  They generate the
keyword query distribution from one year's Ubuntu IRC logs as the auxiliary
information to infer the encrypted keywords in the logs of a later year.  
\item 
Pouliot {\em et al.} \cite{pouliot16} consider the inference attacks against
the keywords in the Enron email dataset \cite{klimt04}.  They first partition
each user's emails into two non-overlapping sets (i.e., training and testing
sets).  They then generate the necessary auxiliary information from the
training set, and infer the content of the testing set that is encrypted. 
\item 
Other studies \cite{zhang16b,cash15,islam12} also leverage the Enron
email dataset.  They create a Zipfian synthetic keyword query distribution
from the keyword list of the whole Enron dataset as the auxiliary information,
and use it to infer the keywords in the original dataset that is encrypted. 
\end{itemize}

We observe that previous studies mainly obtain the auxiliary information from
{\em private}
\cite{grubbs16,pouliot16,zhang16b,cash15,islam12,bindschaedler18} or 
{\em public} \cite{grubbs17,naveed15,bindschaedler18} sources. By private, we
mean that the auxiliary information is originally protected but is obtained
through unintended data releases \cite{arrington06}, data breaches
\cite{hackett16}, or stolen storage devices \cite{hddtheft}. By public, we
mean that the auxiliary information can be legitimately accessed by the
adversary.    

In this work, we mainly focus on the private auxiliary information, which we
model as the {\em plaintext chunks of a prior (non-latest) backup}.
Nevertheless, in our evaluation (see Section~\ref{sec:evaluation}), we also
address the public auxiliary information, in which we choose a virtual disk
image that is publicly accessible. 

We note that the success of frequency analysis heavily depends on how accurate
the available auxiliary information describes the backups.
Our focus is {\em not} to address how to obtain accurate auxiliary
information, which we pose as future work; instead, given the available
auxiliary information, we study how an adversary can design severe attacks
based on frequency analysis and how we can defend against the attacks.  

\subsection{Attack Modes}

Based on the available auxiliary information (which describes a prior backup),
the primary goal of the adversary is to {\em infer} the content of the
plaintext chunks that are mapped to the ciphertext chunks of the latest
backup.  The attack can be based on two modes:
\begin{itemize}[leftmargin=*]
\item
\emph{Ciphertext-only mode:} It models a typical case in which the adversary
can access the ciphertext chunks of the latest backup (as well as the
auxiliary information about a prior backup). 
\item
\emph{Known-plaintext mode:} It models a more severe case in which a powerful
adversary not only can access the ciphertext chunks of the latest backup and
the auxiliary information about a prior backup as in ciphertext-only mode, but
also knows a small fraction of the ciphertext-plaintext chunk pairs about the
latest backup (e.g., from stolen devices \cite{darrow15}). 
\end{itemize}

In both attack modes, we make the following assumptions on the capabilities of
an adversary.
\begin{itemize}[leftmargin=*]
\item 
The adversary can monitor the processing sequence of the storage system and
access the {\em logical order} of ciphertext chunks of the latest backup
before deduplication.  Our rationale is that existing deduplicated storage
systems \cite{zhu08,xia11} often process chunks in logical order, so as to
effectively cache metadata for efficient deduplication. 
\item 
The adversary cannot access any metadata information (e.g., the fingerprint
index, file recipes, key recipes of all files). In practice, we do not apply
deduplication to the metadata, which can be protected by conventional
encryption.  For example, the file recipes and key recipes can be encrypted by
user-specific secret keys.
\item
The adversary cannot identify which prior backup to which a stored ciphertext
chunk belongs by analyzing the physical storage space, as the storage system
can store ciphertext chunks in randomized physical addresses or in commercial
public clouds (the latter is more difficult to access directly).
\end{itemize}

\subsection{Other Attacks}

While this work focuses on frequency analysis, another inference attack based
on combinatorial optimization, called $l_p$-optimization, has been proposed to
attack deterministic encryption \cite{naveed15}.  Nevertheless, frequency
analysis is shown to be as effective as the $l_p$-optimization attack 
\cite{naveed15}, and later studies \cite{lacharite15,pouliot16} also state
that both frequency analysis and $l_p$-optimization may have equivalent
severity.  
	  
We do not consider other threats against encrypted deduplication, as they can
be addressed independently by existing approaches.  For example, the
side-channel attack against encrypted deduplication \cite{halevi11,harnik10}
can be addressed by server-side deduplication \cite{harnik10,li15} and proof
of ownership \cite{halevi11}; the leakage of access pattern \cite{islam12} can
be addressed by oblivious RAM \cite{shi11} and blind storage \cite{naveed14}.

\section{Attacks}
\label{sec:attack}

\begin{table}[!t]
\caption{Major notation used in this paper.}
\label{tab:notation}
\vspace{-6pt}
\renewcommand{\arraystretch}{1.1}
\begin{small}
\centering
\begin{tabular}{|l|p{4.5in}|}
\hline
{\bf Notation} & {\bf Description} \\
\hline
\multicolumn{2}{|c|}{\bf Defined in Section~\ref{sec:attack}}\\
\hline
$\mathbf{C}$ & sequence of ciphertext chunks $\langle C_1, \ldots \rangle$
in logical order for the latest backup\\
\hline
$\mathbf{M}$ & sequence of plaintext chunks $\langle M_1,\ldots \rangle$ in
logical order for a prior backup\\
\hline
$\mathbf{F_C}$  & associative array that maps each ciphertext chunk in
$\mathbf{C}$ to its frequency\\
\hline
$\mathbf{F_M}$  & associative array that maps each plaintext chunk in
$\mathbf{M}$ to its frequency\\ 
\hline
$\mathcal{T}$ & set of inferred ciphertext-plaintext chunk pairs\\
\hline
$\mathcal{L}_C$ & set of left neighbors of ciphertext chunk $C$ \\ 
\hline
$\mathcal{L}_M$ & set of left neighbors of plaintext chunk $M$  \\
\hline
$\mathcal{R}_C$ & set of right neighbors of ciphertext chunk $C$ \\ 
\hline
$\mathcal{R}_M$ & set of right neighbors of plaintext chunk $M$  \\ 
\hline
$\mathcal{G}$ & set of currently inferred ciphertext-plaintext chunk pairs \\ 
\hline
$u$ & number of ciphertext-plaintext chunk pairs returned from frequency
analysis during the initialization of $\mathcal{G}$ \\
\hline
$v$ & number of ciphertext-plaintext chunk pairs returned from frequency
analysis in each iteration of locality-based attack\\
\hline
$w$ & maximum size of $\mathcal{G}$ \\
\hline
$\mathbf{L_C}$ & associative array that maps each ciphertext chunk in
$\mathbf{C}$ to its left neighbor and co-occurrence frequency\\
\hline
$\mathbf{L_M}$ & associative array that maps each plaintext chunk in
$\mathbf{M}$ to its left neighbor and co-occurrence frequency\\
\hline
$\mathbf{R_C}$ & associative array that maps each ciphertext chunk in
$\mathbf{C}$ to its right neighbor and co-occurrence frequency\\
\hline
$\mathbf{R_M}$ & associative array that maps each plaintext chunk in
$\mathbf{M}$ to its right neighbor and co-occurrence frequency\\
\hline
\multicolumn{2}{|c|}{\bf Defined in Section~\ref{sec:defense}} \\
\hline
$K_S$ & segment-based key of segment $S$ \\
\hline
$h$ & minimum fingerprint of chunks in a segment \\
\hline
\end{tabular}
\end{small}
\end{table}

We present inference attacks based on frequency analysis against encrypted
deduplication.  We first present a basic attack, which builds on classical
frequency analysis to infer plaintext content in encrypted deduplication.  We
next propose a more severe locality-based attack, which enhances the basic
attack by exploiting chunk locality.  Furthermore, we combine the
locality-based attack with the chunk size information, and propose an advanced
locality-based attack against variable-size chunks.  

Table~\ref{tab:notation} summarizes the major notation used in this paper. We
first formalize the adversarial goal of our proposed attacks
based on the threat model in Section~\ref{sec:threat}.
Let $\mathbf{C} = \langle C_1, C_2, \ldots \rangle$ be the sequence of
ciphertext chunks in logical order for the latest backup, and $\mathbf{M} =
\langle M_1, M_2, \ldots \rangle$ be the sequence of plaintext chunks in
logical order for a prior backup (i.e., $\mathbf{M}$ is the auxiliary
information).  Both $\mathbf{C}$ and $\mathbf{M}$ show the logical orders
of chunks before deduplication as perceived by the adversary (i.e., identical
chunks may repeat), and each of them can have multiple identical
chunks that have the same content.  Note that both
$\mathbf{C}$ and $\mathbf{M}$ do not necessarily have the same number of
chunks.  Furthermore, the $i$-th plaintext chunk $M_i$ in $\mathbf{M}$ (where
$i\ge 1$) is not necessarily mapped to the $i$-th ciphertext chunk in
$\mathbf{C}$; in fact, $M_i$ may not be mapped to any ciphertext chunk in
$\mathbf{C}$, for example, when $M_i$ has been updated before the latest
backup is generated.  Given $\mathbf{C}$ and $\mathbf{M}$, the goal of an
adversary is to infer the content of the original plaintext chunks in
$\mathbf{C}$. 

We quantify the severity of an attack using the {\em inference rate}, defined
as the ratio of the number of unique ciphertext chunks whose plaintext chunks
are successfully inferred over the total number of unique ciphertext chunks in
the latest backup; a higher inference rate implies that the attack is more
severe.  

\subsection{Basic Attack}
\label{subsec:basic}

We first demonstrate how we can apply frequency analysis to infer the original
plaintext chunks of the latest backup in encrypted deduplication.  We call
this attack the {\em basic attack}. Note that the basic attack
is ineffective in inference (see below and the evaluation in
Section~\ref{sec:evaluation}), yet we use it to guide our design of the
locality-based attack in Section~\ref{subsec:locality}.

\paragraph{Overview:} In the basic attack, we identify each chunk by its
fingerprint, and count the frequency of each chunk by the number of
fingerprints that appear in a backup.  Thus, a chunk (or a fingerprint) has a
high frequency if there exist many identical chunks with the same content.  We
sort the chunks of both $\mathbf{C}$ and $\mathbf{M}$ by their frequencies,
and infer that the $i$-th frequent plaintext chunk in $\mathbf{M}$ is the
original plaintext chunk of the $i$-th frequent ciphertext chunk in
$\mathbf{C}$.  Our rationale is that the frequency of a plaintext chunk is
correlated to that of the corresponding ciphertext chunk due to deterministic
encryption. 

\paragraph{Algorithm details:} Algorithm~\ref{alg:basic} shows the pseudo-code
of the basic attack.  It takes $\mathbf{C}$ and $\mathbf{M}$ as input, and
returns the result set $\mathcal{T}$ of all inferred ciphertext-plaintext
chunk pairs. It first calls the function {\sc Count} to obtain the frequencies
of all ciphertext and plaintext chunks, identified by fingerprints, in
associative arrays $\mathbf{F_C}$ and $\mathbf{F_M}$, respectively
(Lines~2-3).  It then calls the function {\sc Freq-analysis} to infer the set
$\mathcal{T}$ of ciphertext-plaintext chunk pairs (Line~4), and returns
$\mathcal{T}$ (Line~5).

\begin{algorithm}[t]
\caption{Basic Attack}
\label{alg:basic}
\begin{small}
\begin{algorithmic}[1]
\Procedure{Basic attack}{$\mathbf{C}, \mathbf{M}$}		
	\State $\mathbf{F_C}\gets${\sc Count}$(\mathbf{C})$
	\State $\mathbf{F_M}\gets${\sc Count}$(\mathbf{M})$
	\State $\mathcal{T}\gets${\sc Freq-analysis}$(\mathbf{F_C}, \mathbf{F_M})$
	\State \Return $\mathcal{T}$
\EndProcedure
\Statex
\Function{Count}{$\mathbf{X}$}
\State Initialize $\mathbf{F_X}$
\For{each $X$ in $\mathbf{X}$}
	\If{$X$ does not exist in $\mathbf{F_X}$}
	  \State Initialize $\mathbf{F_X}[X] \gets 0$
	\EndIf
	\State $\mathbf{F_X}[X]\gets\mathbf{F_X}[X] + 1$
\EndFor
\State\Return $\mathbf{F_X}$
\EndFunction
\Statex
\Function{Freq-analysis}{$\mathbf{F_C}, \mathbf{F_M}$}
	\State Sort $\mathbf{F_C}$ by frequency
	\State Sort $\mathbf{F_M}$ by frequency
	\State $min\gets\min\{|\mathbf{F_C}|, |\mathbf{F_M}|\}$
\For{$i=1$ to $min$}
	\State $C\gets$ $i$-th frequent ciphertext chunk
	\State $M\gets$ $i$-th frequent plaintext chunk
	\State Add $(C, M)$ to $\mathcal{T}'$
\EndFor
	\State \Return $\mathcal{T}'$
\EndFunction
\end{algorithmic}
\end{small}
\end{algorithm}

The function {\sc Count} constructs an associative array $\mathbf{F_X}$
(where $\mathbf{X}$ can be either $\mathbf{C}$ and $\mathbf{M}$) that
holds the frequencies of all chunks.  If a chunk $X$ does not exist in
$\mathbf{F_X}$ (i.e., its fingerprint is not found), then the function adds
$X$ to $\mathbf{F_X}$ and initializes $\mathbf{F_X}[X]$ as zero (Lines~10-12). 
The function then increments $\mathbf{F_X}[X]$ by one (Line~13).  

The function {\sc Freq-analysis} performs frequency analysis based on 
$\mathbf{F_C}$ and $\mathbf{F_M}$.  It first
sorts each of $\mathbf{F_C}$ and $\mathbf{F_M}$ by frequency (Lines~18-19).
Since $\mathbf{F_C}$ and $\mathbf{F_M}$ may not have the same number of
elements, it finds the minimum number of elements in $\mathbf{F_C}$ and
$\mathbf{F_M}$ (Line~20).  Finally, it returns the ciphertext-plaintext
chunk pairs, in which both the ciphertext and plaintext chunks of each pair
have the same rank (Lines~21-26). 
		  
\paragraph{Discussion:}  The basic attack demonstrates how frequency analysis
can be applied to encrypted deduplication.  However, it only achieves a small
inference rate, as shown in our trace-driven evaluation (see
Section~\ref{sec:evaluation}).  One reason is that the basic attack is
sensitive to data updates that occur across different versions of backups over
time.  An update to a chunk can change the frequency ranks of multiple chunks,
including the chunk itself and other chunks with similar frequencies.   
Specifically, an update to a ciphertext chunk $C$ in the latest backup can
lower the frequency rank of $C$, which now has fewer copies of identical
ciphertext chunks with the same content, while promoting the ranks of other
ciphertext chunks whose ranks are just below $C$.
Another reason is that there exist many ties, in which chunks have the same
frequency.  How to break a tie during sorting also affects the frequency rank
and hence the inference results of the tied chunks.  In the following, we
extend the basic attack to improve its inference rate. 

\subsection{Locality-based Attack}
\label{subsec:locality}

The \emph{locality-based attack} exploits \emph{chunk locality} to make
frequency analysis more effective. 

\paragraph{Overview:}  We first define the notation that captures the notion
of chunk locality.  Consider two ordered pairs $\langle C_i, C_{i+1} \rangle$
and $\langle M_i, M_{i+1} \rangle$ of neighboring ciphertext and plaintext
chunks in $\mathbf{C}$ and $\mathbf{M}$, respectively.  We say that $C_i$ is
the {\em left} neighbor of $C_{i+1}$, while $C_{i+1}$ is the {\em right}
neighbor of $C_i$; similar definitions apply to $M_{i}$ and $M_{i+1}$.
Note that a ciphertext chunk in $\mathbf{C}$ or a plaintext chunk in
$\mathbf{M}$ may repeat many times (i.e., there are many duplicate copies), so
if we identify each chunk by its fingerprint, it can be associated with more
than one left or right neighbor. 
Let $\mathcal{L}_C$ and $\mathcal{R}_C$ be the sets of left neighbors and
right neighbors of a ciphertext chunk $C$, respectively, and $\mathcal{L}_M$
and $\mathcal{R}_M$ be the left and right neighbors of a plaintext chunk $M$,
respectively. 

Our insight is that if a plaintext chunk $M$ of a prior backup has been
identified as the original plaintext chunk of a ciphertext chunk $C$ of the
latest backup, then the left and right neighbors of $M$ are also likely to be
original plaintext chunks of the left and right neighbors of $C$, mainly
because chunk locality implies that the ordering of chunks is likely to be
preserved across backups. 
In other words, for any inferred ciphertext-plaintext chunk pair $(C, M)$,
we further infer more ciphertext-plaintext chunk pairs through the left and
right neighboring chunks of $C$ and $M$, and repeat the same inference on
those newly inferred chunk pairs.  Thus, we can significantly increase the
attack severity. 

The locality-based attack operates on an {\em inferred set} $\mathcal{G}$,
which stores the currently inferred set of ciphertext-plaintext chunks pairs.
How to initialize $\mathcal{G}$ depends on the attack modes (see
Section~\ref{sec:threat}).  In ciphertext-only mode, in which an adversary
only knows $\mathbf{C}$ and $\mathbf{M}$, we apply frequency analysis to find
the most frequent ciphertext-plaintext chunk pairs and add them to
$\mathcal{G}$.  Here, we configure a parameter $u$ 
(e.g., $u=$~1 by default in our implementation)
to indicate the number of most frequent chunk pairs to be returned.
Our rationale is that the top-frequent chunks have significantly higher
frequencies (see Figure~\ref{fig:freq-dist}) than the other chunks, and their
frequency ranks are stable across different backups.  This ensures the
correctness of the ciphertext-plaintext chunk pairs in $\mathcal{G}$ with a
high probability throughout the attack. On the other hand, in known-plaintext
mode, in which the adversary knows some leaked ciphertext-plaintext chunk
pairs about $\mathbf{C}$ for the latest backup, we initialize $\mathcal{G}$
with the set of leaked chunk pairs that are also in $\mathbf{M}$. 

The locality-based attack proceeds as follows. In each iteration, it picks one
ciphertext-plaintext chunk pair $(C, M)$ from $\mathcal{G}$.  It collects the
corresponding sets of neighboring chunks $\mathcal{L}_C$, $\mathcal{L}_M$,
$\mathcal{R}_C$, and $\mathcal{R}_M$.  We apply frequency analysis to find the
most frequent ciphertext-plaintext chunk pairs from each of $\mathcal{L}_C$ and
$\mathcal{L}_M$, and similarly from $\mathcal{R}_C$ and $\mathcal{R}_M$. In
other words, we find the left and right neighboring chunks of $C$ and $M$ that
have the most {\em co-occurrences} with $C$ and $M$ themselves, respectively. 
We configure a parameter $v$ (e.g., $v=$~15 by default in our
implementation) to indicate the number of most frequent chunk pairs returned
from the frequent analysis algorithm being called in each iteration.  
A larger $v$ increases the number of inferred ciphertext-plaintext chunk
pairs, but it also potentially compromises the inference accuracy.  The attack
adds all inferred chunk pairs into $\mathcal{G}$, and iterates until all
inferred chunk pairs in $\mathcal{G}$ have been processed. 

Note that $\mathcal{G}$ may grow very large as the backup size increases. 
A very large $\mathcal{G}$ can exhaust memory space.  We
configure a parameter $w$ 
(e.g., $w=$~200,000 by default in our implementation) 
to bound the maximum size of $\mathcal{G}$. 

In our evaluation (see Section~\ref{sec:evaluation}), we carefully examine the
impact of the configurable parameters $u$, $v$, and $w$. 

\begin{algorithm}[!t]
\caption{Locality-based Attack} 
\label{alg:locality-attack}
\begin{small}
\setlength\multicolsep{3pt}
\begin{multicols}{2}
\begin{algorithmic}[1]
\Procedure{Locality attack}{$\mathbf{C}, \mathbf{M}, u, v, w$} 		
\State $(\mathbf{F_C}, \mathbf{L_C}, \mathbf{R_C}) \gets$
{\sc Count}$(\mathbf{C})$ 
\State $(\mathbf{F_M}, \mathbf{L_M}, \mathbf{R_M}) \gets$
{\sc Count}$(\mathbf{M})$ 
\If{ciphertext-only mode}
  \State $\mathcal{G} \gets\Call{Freq-analysis}{\mathbf{F_C}, \mathbf{F_M}, u}$
\ElsIf{known-plaintext mode}
  \State $\mathcal{G} \gets$ set of leaked
	ciphertext-plaintext chunk pairs that appear in both $\mathbf{C}$ and
	$\mathbf{M}$
\EndIf
\State $\mathcal{T}\gets\mathcal{G}$
\While{$\mathcal{G}$ is non-empty} 
  \State Remove $(C, M)$ from $\mathcal{G}$ 
  \State $\mathcal{T}_l\gets$
    $\Call{Freq-analysis}{
	  \mathbf{L_C}[C], \mathbf{L_M}[M], v}$

  \State $\mathcal{T}_r\gets$
            $\Call{Freq-analysis}{
	  \mathbf{R_C}[C], \mathbf{R_M}[M], v}$
  \For{each $(C, M)$ in $\mathcal{T}_l\cup\mathcal{T}_r$}
	 \If{$(C,*)$ is not in $\mathcal{T}$}
        \State Add $(C, M)$ to $\mathcal{T}$
		\If{$|\mathcal{G}| \le w$ (i.e., $\mathcal{G}$ is not full)}
           \State Add $(C, M)$ to $\mathcal{G}$
		\EndIf
	 \EndIf
  \EndFor
\EndWhile 
\State \Return $\mathcal{T}$ 
\EndProcedure
\Statex
\Function{Count}{$\mathbf{X}$} 		
\State Initialize $\mathbf{F_X}$, $\mathbf{L_X}$, and $\mathbf{R_X}$
\For{each $X$ in $\mathbf{X}$}
    \If{$X$ does not exist in $\mathbf{F_X}$}
	   \State Initialize $\mathbf{F_X}[X] \gets 0$
    \EndIf
	\State $\mathbf{F_X}[X] \gets \mathbf{F_X}[X] + 1$
	\If{$X$ has a left neighbor $X_l$}
	   \If{$X_l$ does not exist in $\mathbf{L_X}[X]$}
		  \State Initialize $\mathbf{L_X}[X][X_l] \gets 0$
	   \EndIf
	   \State $\mathbf{L_X}[X][X_l] \gets \mathbf{L_X}[X][X_l] + 1$
	\EndIf
	\If{$X$ has a right neighbor $X_r$}
	   \If{$X_r$ does not exist in $\mathbf{R_X}[X]$}
		  \State Initialize $\mathbf{R_X}[X][X_r] \gets 0$
	   \EndIf
	   \State $\mathbf{R_X}[X][X_r] \gets \mathbf{R_X}[X][X_r] + 1$
	\EndIf
\EndFor
\State \Return $(\mathbf{F_X}, \mathbf{L_X}, \mathbf{R_X})$ 
\EndFunction 
\Statex
\Function{Freq-analysis}{$\mathbf{Y_C}$, $\mathbf{Y_M}, x$} 		
	\State Sort $\mathbf{Y_C}$ by frequency
	\State Sort $\mathbf{Y_M}$ by frequency
    \For{$i=1$ to $\min\{x, |\mathbf{Y_C}|, |\mathbf{Y_M}|\}$}
 	  \State $C\gets$ $i$-th frequent ciphertext chunk
	  \State $M\gets$ $i$-th frequent plaintext chunk
	  \State Add $(C, M)$ to $\mathcal{T}'$
    \EndFor
	\State \Return $\mathcal{T}'$
\EndFunction 
\end{algorithmic}
\end{multicols}
\end{small}
\end{algorithm}

\paragraph{Algorithm details:}
Algorithm~\ref{alg:locality-attack} shows the pseudo-code of the
locality-based attack.  It takes $\mathbf{C}$, $\mathbf{M}$, $u$, $v$, and $w$
as input, and returns the result set $\mathcal{T}$ of all inferred
ciphertext-plaintext chunk pairs.  It first calls the function {\sc Count} to
obtain the following associative arrays: $\mathbf{F_C}$, which stores the
frequencies of all ciphertext chunks, as well as $\mathbf{L_C}$ and
$\mathbf{R_C}$, which store the co-occurrence frequencies of the left and
right neighbors of all ciphertext chunks, respectively (Line~2); similarly, it
obtains the associative arrays $\mathbf{F_M}$, $\mathbf{L_M}$, and
$\mathbf{R_M}$ for the plaintext chunks (Line~3). 
It then initializes the inferred set $\mathcal{G}$, either by obtaining $u$
most frequent ciphertext-plaintext chunk pairs from frequency
analysis in ciphertext-only mode, or by adding the set of leaked
ciphertext-plaintext chunk pairs that appear in both the latest and prior
backups (i.e., $\mathbf{C}$ and $\mathbf{M}$, respectively) in known-plaintext
mode (Lines~4-8).  It also initializes $\mathcal{T}$ with $\mathcal{G}$
(Line~9). 

In the main loop (Lines~10-22), the algorithm removes a pair $(C, M)$ from
$\mathcal{G}$ (Line~11) and uses it to infer additional ciphertext-plaintext
chunk pairs from the neighboring chunks of $C$ and $M$.  It first examines all
left neighbors by running the function {\sc Freq-analysis} on
$\mathbf{L_C}[C]$ and $\mathbf{L_M}[M]$, and stores $v$ most frequent
ciphertext-plaintext chunk pairs in $\mathcal{T}_l$ (Line~12).
Similarly, it examines all right neighbors and stores the results in
$\mathcal{T}_r$ (Line~13).  
For each $(C, M)$ in $\mathcal{T}_l \cup \mathcal{T}_r$, if $(C,*)$ is not in
$\mathcal{T}$ (i.e., the original plaintext chunk of $C$ has not been inferred
yet), we add $(C,M)$ to $\mathcal{T}$ and also to $\mathcal{G}$ if
$\mathcal{G}$ is not full (Lines~14-21).  The main loop iterates until
$\mathcal{G}$ becomes empty.  Finally, $\mathcal{T}$ is returned. 

Both the functions {\sc Count} and {\sc Freq-analysis} are similar to those in
the basic attack (see Algorithm~\ref{alg:basic}), with the following
extensions.  For {\sc Count}, in addition to constructing the associative array
$\mathbf{F_X}$ (where $\mathbf{X}$ can be either $\mathbf{C}$ and
$\mathbf{M}$) that holds the frequencies of all chunks, it also constructs the
associative arrays $\mathbf{L_X}$ and $\mathbf{R_X}$ that hold the
co-occurrence frequencies
of the left and right neighbors of each chunk $X$, respectively.  For 
{\sc Freq-analysis}, it now performs frequency analysis on the associative
arrays $\mathbf{Y_C}$ and $\mathbf{Y_M}$, in which $\mathbf{Y_C}$ (resp.
$\mathbf{Y_M}$) refers to either $\mathbf{F_C}$ (resp. $\mathbf{F_M}$) that
holds the frequency counts of all chunks, or $\mathbf{L_C}[C]$ and
$\mathbf{R_C}[C]$ (resp.  $\mathbf{L_M}[M]$ and $\mathbf{R_M}[M]$) that hold
the frequency counts of all ordered pairs of chunks associated with ciphertext
chunk $C$ (resp. plaintext chunk $M$).  Also, {\sc Freq-analysis} only returns
$x$ (where $x$ can be either $u$ or $v$) most frequent ciphertext-plaintext
chunk pairs. 
	
\begin{figure}[!t]
\centering
\includegraphics[width=3.5in]{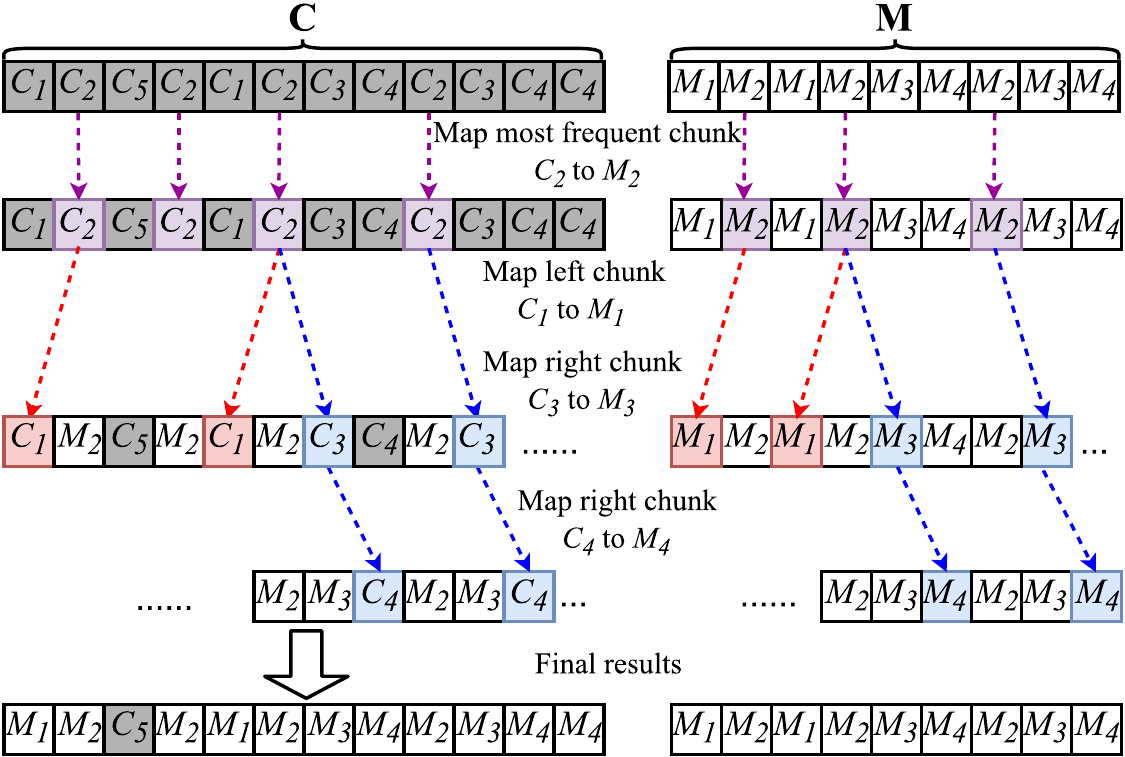}
\caption{Example of the locality-based attack.}
\label{fig:example}
\end{figure}

\paragraph{Example:} Figure~\ref{fig:example} shows an example of how the
locality-based attack works.  Here, we consider ciphertext-only mode.
Suppose that we have obtained the auxiliary information
$\mathbf{M}=$ $\langle M_1, M_2, M_1, M_2, M_3, M_4, M_2,$ $M_3, M_4\rangle$ of 
some prior backup, and use it to infer the original plaintext chunks of 
$\mathbf{C}=$ $\langle C_1, C_2, C_5, C_2, C_1,$ $C_2, C_3,$ $C_4, C_2, C_3, C_4, C_4\rangle$ of the latest backup.  We set $u = v = 1$, and
$w\rightarrow\infty$ (i.e., the inferred set $\mathcal{G}$ is unbounded). 
We assume that the ground truth is that the original plaintext chunk of the
ciphertext chunk $C_i$ is $M_i$ for $i=1, 2, 3, 4$, while that of $C_5$ is
some new plaintext chunk not in $\mathbf{M}$ (note that in reality, an
adversary does not know the ground truth). 

We first apply frequency analysis and find that $(C_2,M_2)$ is the most
frequent ciphertext-plaintext chunk pair, so we initialize
$\mathcal{G}=\{(C_2, M_2)\}$ and add it into $\mathcal{T}$.  We then remove and
operate on $(C_2, M_2)$ from $\mathcal{G}$, and find that
$\mathcal{L}_{C_2}=\{C_1, C_4, C_5\}$, $\mathcal{L}_{M_2} = \{M_1, M_4\}$,
$\mathcal{R}_{C_2} = \{C_1, C_3, C_5\}$, and 
$\mathcal{R}_{M_2} = \{M_1, M_3\}$.  From
$\mathcal{L}_{C_2}$ and $\mathcal{L}_{M_2}$, we find that $(C_1,M_1)$ is the
most frequent ciphertext-plaintext chunk pair, while from $\mathcal{R}_{C_2}$
and $\mathcal{R}_{M_2}$, we find $(C_3,M_3)$.  Thus, we add both $(C_1, M_1)$
and $(C_3, M_3)$ into $\mathcal{G}$ and $\mathcal{T}$.  We repeat the
processing on $(C_1, M_1)$ and $(C_3, M_3)$, and we can infer another pair
$(C_4, M_4)$ from the right neighbors of $(C_3, M_3)$. 

To summarize, the locality-based attack can successfully infer the original
plaintext chunks of all four ciphertext chunks $C_1$, $C_2$, $C_3$, and $C_4$.
It cannot infer the original plaintext chunk of $C_5$, as it does not appear
in $\mathbf{M}$. 

\subsection{Advanced Locality-based Attack}

Based on the framework of the locality-based attack, we propose an advanced
locality-based attack that specifically targets variable-size chunks generated
from content-defined chunking (see Section~\ref{subsec:dedup}).  Specifically,
if the generated chunks have varying sizes, the adversary can observe the size
of each ciphertext chunk before deduplication and leverage the size
information to increase the severity of the locality-based attack. 
	
\paragraph{Overview:}
The advanced locality-based attack builds on the observation that if a
ciphertext chunk $C$ corresponds to a plaintext chunk $M$, then the actual
size of $C$ approximates that of $M$.  Suppose that the symmetric encryption
algorithm used by the encrypted deduplication system is based on block ciphers
(e.g., AES),  then both $C$ and $M$ should have the same number of blocks
(i.e., the basic units of block ciphers).  We exploit this additional
information in frequency analysis. 
Specifically, we first classify the sets of ciphertext chunks (i.e.,
$\mathbf{C}$, $\mathcal{L}_C$, $\mathcal{R}_C$) and plaintext chunks (i.e., 
$\mathbf{M}$, $\mathcal{L}_M$, and $\mathcal{R}_M$) by their sizes, measured
in terms of the number of blocks.  For each available chunk size, we relate
top-frequent ciphertext chunks with the top-frequent plaintext chunks that
have the same size.  This improves the accuracy of each inferred
ciphertext-plaintext pair, and hence the inferred neighbors in the iterated
inference of the locality-based attack.    

\paragraph{Algorithm details:}
The advanced locality-based attack extends the original locality-based attack
in Algorithm~\ref{alg:locality-attack} and modifies the function 
{\sc Freq-analysis} (called in Line~5 and Lines~12-13 in
Algorithm~\ref{alg:locality-attack}) to augment frequency analysis with the
knowledge of chunk sizes. 

Algorithm~\ref{alg:advanced} shows the pseudo-code of frequency analysis in
the advanced locality-based attack. As in Algorithm~\ref{alg:locality-attack},
the function {\sc Freq-analysis} takes the associative arrays $\mathbf{Y_C}$
and $\mathbf{Y_M}$, as well as the parameter $x$, as input. It calls the
function {\sc Classify} to classify the ciphertext and plaintext chunks in
$\mathbf{Y_C}$ and $\mathbf{Y_M}$ into $\mathbf{B_C}$ and $\mathbf{B_M}$,
respectively (Lines~2-3), where $\mathbf{B_C}$ (resp. $\mathbf{B_M}$) maps the
ciphertext (resp. plaintext) chunks that have the same sizes to corresponding
frequencies.  It infers $x$ top-frequent ciphertext-plaintext pairs for each
available $s$ (Lines~4-12), and finally returns the inference results
(Line~13).             

The function {\sc Classify} groups the chunks in $\mathbf{Y_X}$ (where
$\mathbf{X}$ can be either $\mathbf{C}$ or $\mathbf{M}$) by their sizes. In
this work, we assume that AES encryption is used and the block size is
16~bytes.  Thus, {\sc Classify} derives the number of blocks $s$ of each
ciphertext or plaintext chunk (denoted by $X$) (Line~18), and stores the
frequency of $X$ in $\mathbf{B_X}[s][X]$ (Line~22).    

\begin{algorithm}[t]
\caption{Frequency Analysis in the Advanced Locality-based Attack}
\label{alg:advanced}
\small
\begin{algorithmic}[1]
\Function{Freq-analysis}{$\mathbf{Y_C}, \mathbf{Y_M}, x$}
\State $\mathbf{B_C} \gets \Call{Classify}{\mathbf{Y_C}}$ 
\State $\mathbf{B_M} \gets \Call{Classify}{\mathbf{Y_M}}$ 
\For{each available size $s$}
  \State Sort $\mathbf{B_C}[s]$ by frequency
  \State Sort $\mathbf{B_M}[s]$ by frequency
  \For{$i=1$ to $x$}
    \State $C\gets$ $i$-th frequent ciphertext chunk in $\mathbf{B_C}[s]$
    \State $M\gets$ $i$-th frequent plaintext chunk in $\mathbf{B_M}[s]$
    \State Add $(C, M)$ to $\mathcal{T}'$
  \EndFor
\EndFor
\State\Return $\mathcal{T}'$
\EndFunction
\Statex
\Function{Classify}{$\mathbf{Y_X}$}
\State Initialize $\mathbf{B_X}$ 
\For{each $X$ in $Y_X$}
  \State $s \gets \lceil \frac{\text{size}(X)}{16} \rceil$
  \If{$s$ does not exist in $\mathbf{B_X}$}
    \State Initialize $\mathbf{B_X}[s] \gets 0$
  \EndIf
  \State $\mathbf{B_X}[s][X] \gets \mathbf{Y_X}[X]$
\EndFor
\State\Return $\mathbf{B_X}$
\EndFunction
\end{algorithmic}
\end{algorithm}

\section{Attack Evaluation}
\label{sec:evaluation}

We present trace-driven evaluation results on the severity of frequency
analysis against encrypted deduplication. 

\subsection{Datasets}
\label{subsec:dataset}

We consider three datasets in our evaluation. 

\paragraph{FSL:} This dataset is collected by the File systems and Storage Lab
(FSL) at Stony Brook University \cite{FSL14,sun16,tarasov12} and describes
real-world storage patterns.  We focus on the \emph{Fslhomes} dataset, which
contains the daily snapshots of users' home directories on a shared file
system.  Each snapshot is represented by a collection of 48-bit chunk
fingerprints produced by variable-size chunking of different average sizes.
We pick the snapshots from January 22 to May 21 in 2013, and fix the average
size as 8\,KB for our evaluation.  We select six users (User4, User7, User12,
User13, User15, and User28) that have the complete daily snapshots over the
whole duration.  We aggregate each user's snapshots on a monthly basis (on
January 22, February 22, March 22, April 21, and May 21), and hence form five
monthly full backups for all users.  Our post-processed dataset covers a total
of 2.7\,TB of logical data before deduplication, and the overall deduplication
ratio (i.e., the ratio of the logical data size to the physical data size
after deduplication) is 7.6$\times$.  

\paragraph{Synthetic:} This dataset contains a sequence of synthetic backup
snapshots that are generated based on Lillibridge {\em et al.}'s approach
\cite{lillibridge13}.  Specifically, we create an initial snapshot from a
Ubuntu~14.04 virtual disk image (originally with 1.1\,GB of data) with a total
of 4.3\,GB space.  We create a sequence of snapshots starting from the initial
snapshot, such that each snapshot is created from the previous one by randomly
picking 2\% of files and modifying 2.5\% of their content, and also adding
10\,MB of new data.  Finally, we generate a sequence of ten snapshots, each of
which is treated as a backup.  Based on our choices of parameters, the
resulting storage saving of deduplication is around 90\% (or equivalently, the
deduplication ratio is around 10$\times$, which is typical in real-life backup
workloads \cite{wallace12}).  Note that the initial snapshot is publicly
available.  Later in our evaluation, we study the effectiveness of the attacks
by using initial snapshot as the public auxiliary information.
			
\paragraph{VM:} This dataset is collected by ourselves in a real-world
scenario.  It comprises 156 virtual machine (VM) image snapshots for the
students enrolled in a university programming course in Spring 2014. Each
snapshot is represented by the SHA-1 fingerprints of 4\,KB fixed-size chunks.
We treat the VM image snapshot as a weekly backup of a user, and extract 13
weeks of backups of all users.  We remove all zero-filled chunks that dominate
in VM images \cite{jin09}, and obtain a reduced dataset covering 9.11TB of
data.  The overall deduplication ratio of the VM dataset is 47.6$\times$.
Our prior studies \cite{li15,qin17} have also used the variants of the dataset
for evaluation.  Here, we include this dataset for cross-validation of other
datasets in our attack evaluation. 

\subsection{Methodology}
\label{subsec:implementation}

We implement all three inference attacks by processing and comparing the
chunk fingerprints in our datasets.  We benchmark our current implementation
on a Ubuntu~16.04 Linux machine with an AMD Athlon II X4 640 quad-core
3.0\,GHz CPU and 16\,GB RAM.  
In general, both the memory usage and the processing time of the locality-based
attack increase with the total number of unique chunks in a backup (the actual
performance overhead of the locality-based attack depends on the number of
inferred chunk pairs being processed in each iteration). 
Specifically, the locality-based attack takes around 1.8\,GB memory and
15~hours to process an FSL backup that includes around 30~million unique
chunks, and takes around 600\,MB memory and 10~hours to process a VM
backup that includes around 6~million unique chunks.
In the following, we highlight the implementation details of some data
structures used by the attacks. 

\paragraph{Associative arrays:}  Recall that there are three types of
associative arrays: (i) $\mathbf{F_C}$ and $\mathbf{F_M}$, (ii) $\mathbf{L_C}$
and $\mathbf{L_M}$, and (iii) $\mathbf{R_C}$ and $\mathbf{R_M}$ (the latter
two are only used by the locality-based attack).  We implement them as
key-value stores using LevelDB \cite{leveldb}.   Each key-value store is keyed
by the fingerprint of the ciphertext/plaintext chunk.  For $\mathbf{F_C}$ and 
$\mathbf{F_M}$, each entry stores a frequency count; for $\mathbf{L_C}$,
$\mathbf{L_M}$, $\mathbf{R_C}$, and $\mathbf{R_M}$, each entry stores a
sequential list of the fingerprints of all the left/right neighbors of the
keyed chunk and the co-occurrence frequency counts.   
For the latter, keeping neighboring chunks sequentially simplifies our
implementation, but also increases the search time of a particular neighbor
(which dominates the overall running time); we pose the optimization as future
work. 

\paragraph{Inferred set:} We implement the inferred set $\mathcal{G}$ in the
locality-based attack as a first-in-first-out queue, whose maximum size is
bounded by $w$ (see Section~\ref{subsec:locality}).  Each time we remove the
first ciphertext-plaintext chunk pair from the queue for inferring more chunk
pairs from the neighbors. 

\subsection{Results}
\label{sec:attack_result}

We now present the evaluation results and show the inference rate (defined in
Section~\ref{sec:attack}) of each attack under different settings. 

\subsubsection{Impact of Parameters} 

We first evaluate the impact of parameters on the locality-based attack, in
order to justify our choices of parameters.  Recall that the locality-based
attack is configured with three parameters: $u$, $v$, and $w$, 
where $u$ and $v$ are the numbers of ciphertext-plaintext pairs returned by
frequency analysis during the initialization of $\mathcal{G}$ and each
iteration of the locality-based attack, respectively, and $w$ is the maximum
size of $\mathcal{G}$. Here, we focus on the FSL and VM datasets, and
evaluate the attack in ciphertext-only mode.  For the FSL dataset, we use the
backup on March~22 as the auxiliary information in order to infer the plaintext
chunks of the latest backup on May~21; for the VM dataset, we use the 12th
weekly backup to infer the plaintext chunks of the latest 13th weekly backup.
 
\begin{figure*}[!t]
\centering
\subfigure[Varying $u$]{
\includegraphics[width=.31\textwidth]{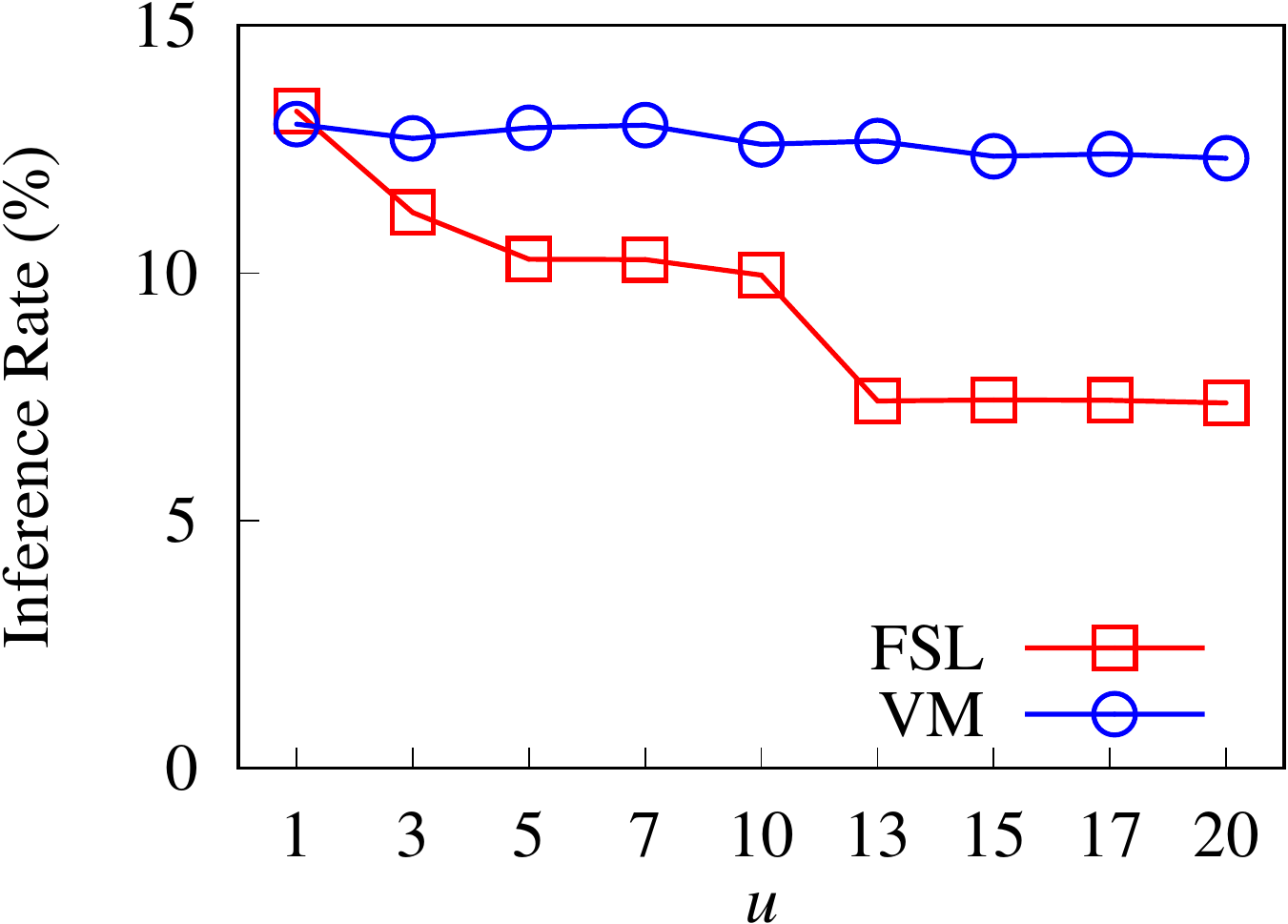}
\label{fig:param-u}
}
\subfigure[Varying $v$]{
\includegraphics[width=.31\textwidth]{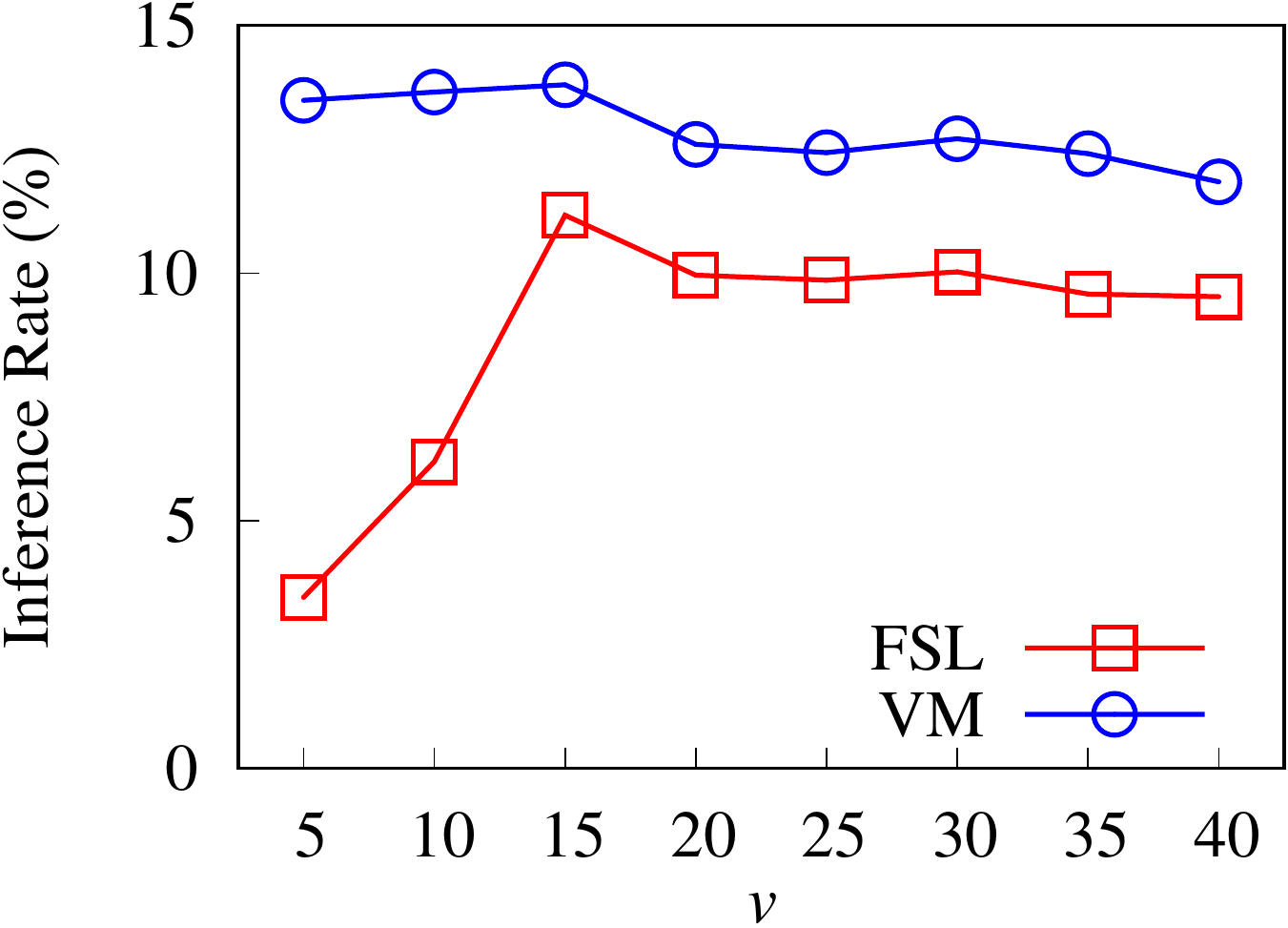}
\label{fig:param-v}
}
\subfigure[Varying $w$]{
\includegraphics[width=.31\textwidth]{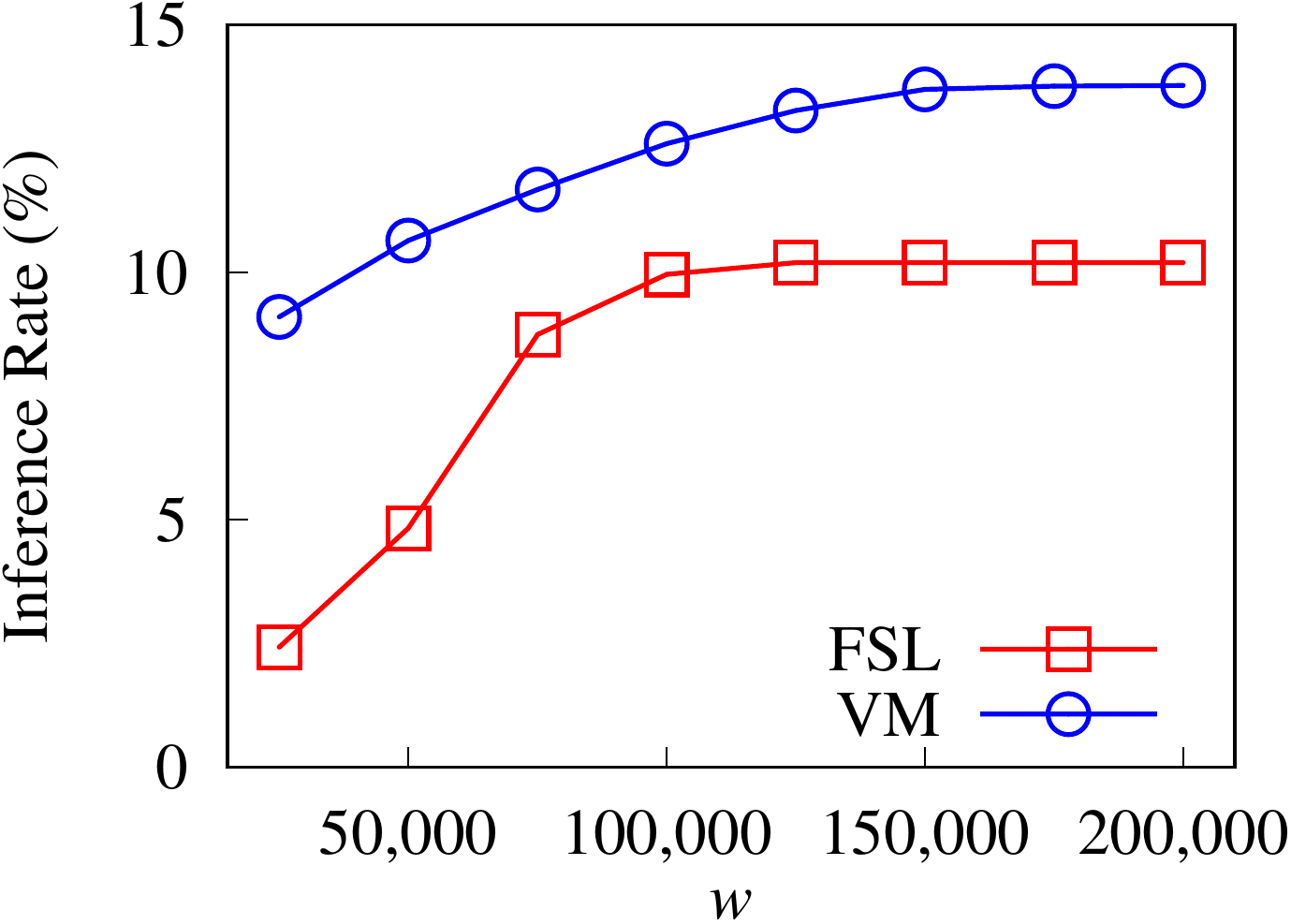}
\label{fig:param-w}
}
\vspace{-6pt}
\caption{Attack evaluation: Impact of parameters on locality-based attack.}
\label{fig:optimal-param}
\end{figure*}

Figure~\ref{fig:param-u} first shows the impact of $u$, in which we fix
$v=$~20 and $w=$~100,000. The inference rate gradually decreases with $u$.
For example, when $u$ increases from 1 to 20, the inference rate decreases
from 13.3\% to 7.4\% and from 13.0\% to 12.3\% for the FSL and VM
datasets, respectively.  A larger $u$ implies that incorrect
ciphertext-plaintext chunk pairs are more likely to be included into the
inferred set during initialization, thereby compromising the inference
accuracy. In addition, the decrease of the inference rate in the VM dataset
is slower than that in the FSL dataset.  The reason is that we use a more
recent VM backup as the auxiliary information and its frequency ranking is
similar to that of the latest backup.

Figure~\ref{fig:param-v} next shows the impact of $v$, in which we fix $u=$~10
and $w=$~100,000. Initially, the inference rate increases with $v$, as the
underlying frequency analysis infers more ciphertext-plaintext chunk pairs in
each iteration.  It hits the maximum value at about 11.2\% (for
the FSL dataset) and 13.8\% (for the VM dataset) when $v=~$15.  When $v$
increases to 40, the inference rate drops slightly to about 9.5\%
and 11.8\% for the FSL and VM datasets, respectively. The reason is that some
incorrectly inferred ciphertext-plaintext chunk pairs are also included into
$\mathcal{G}$, which compromises the inference rate.

Figure~\ref{fig:param-w} finally shows the impact of $w$, in which we fix
$u=$~10 and $v=$~20. A larger $w$ increases the inference rate, since
$\mathcal{G}$ can hold more ciphertext-plaintext chunk pairs across
iterations.  We observe that when $w$ increases beyond 200,000, the inference
rate becomes steady at about 10.2\% and 13.8\% for the FSL and VM
datasets, respectively.  

\subsubsection{Inference Rate in Ciphertext-only Mode}  
\label{subsubsec:ciphertext}

We now evaluate the attacks in ciphertext-only mode.  We select $u=$~1,
$v=$~15, and $w=$~200,000 as default parameters to achieve the highest
possible inference rate.

\paragraph{Varying auxiliary backups:}
First, we choose each of the prior backups as the auxiliary information (we
call them {\em auxiliary backups}), and infer the original plaintext chunks
in the latest backup (i.e., the backup on May~21 in the FSL dataset, the 10th
backup in the synthetic dataset, and the 13th backup in the VM dataset).  

\begin{figure*}[!t]
\centering
\subfigure[FSL dataset]{
\includegraphics[height=1.45in]{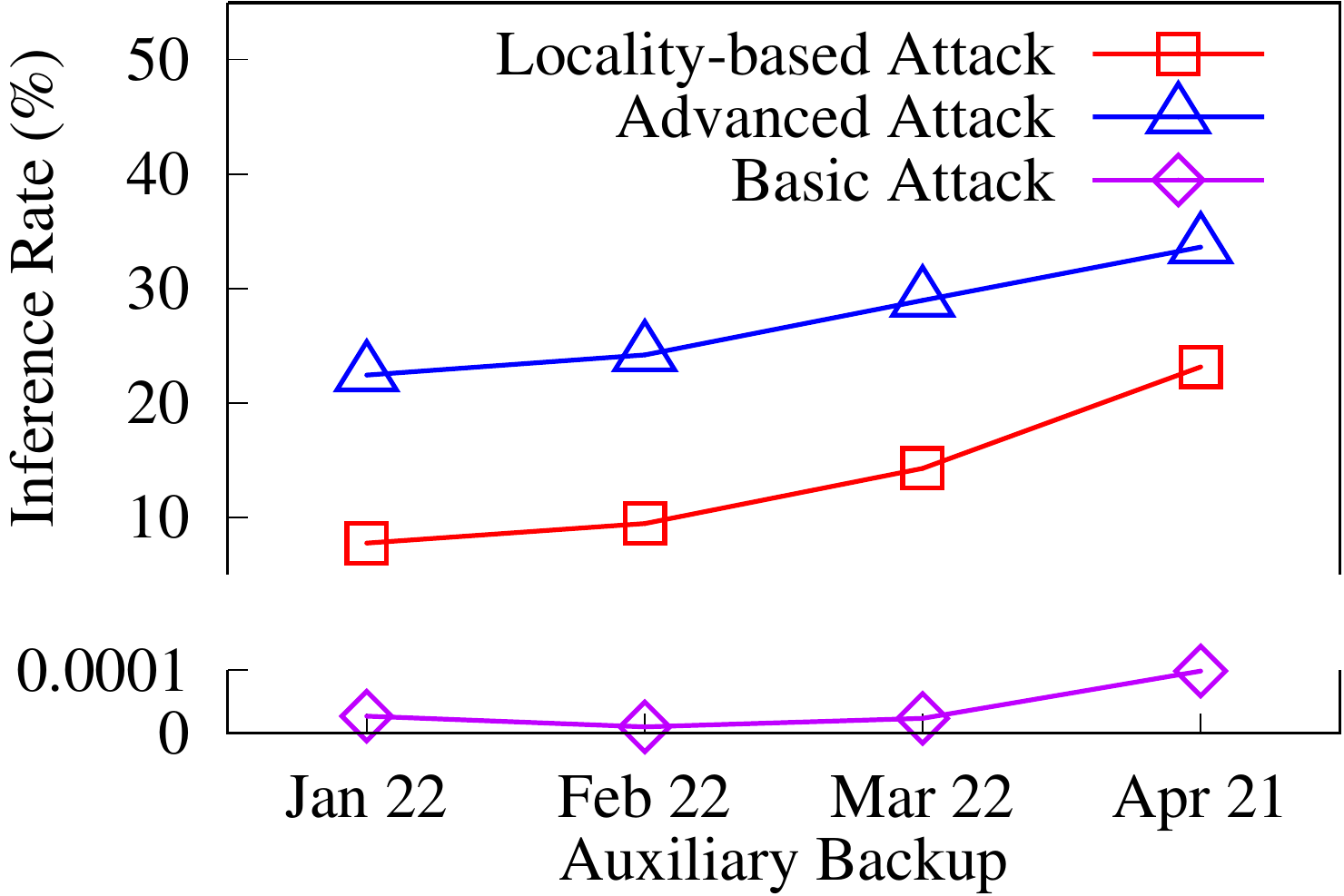}
\label{fig:inference-cipher-only-fsl}
}
\subfigure[Synthetic dataset]{
\label{fig:inference-cipher-only-synthetic}
\includegraphics[height=1.45in]{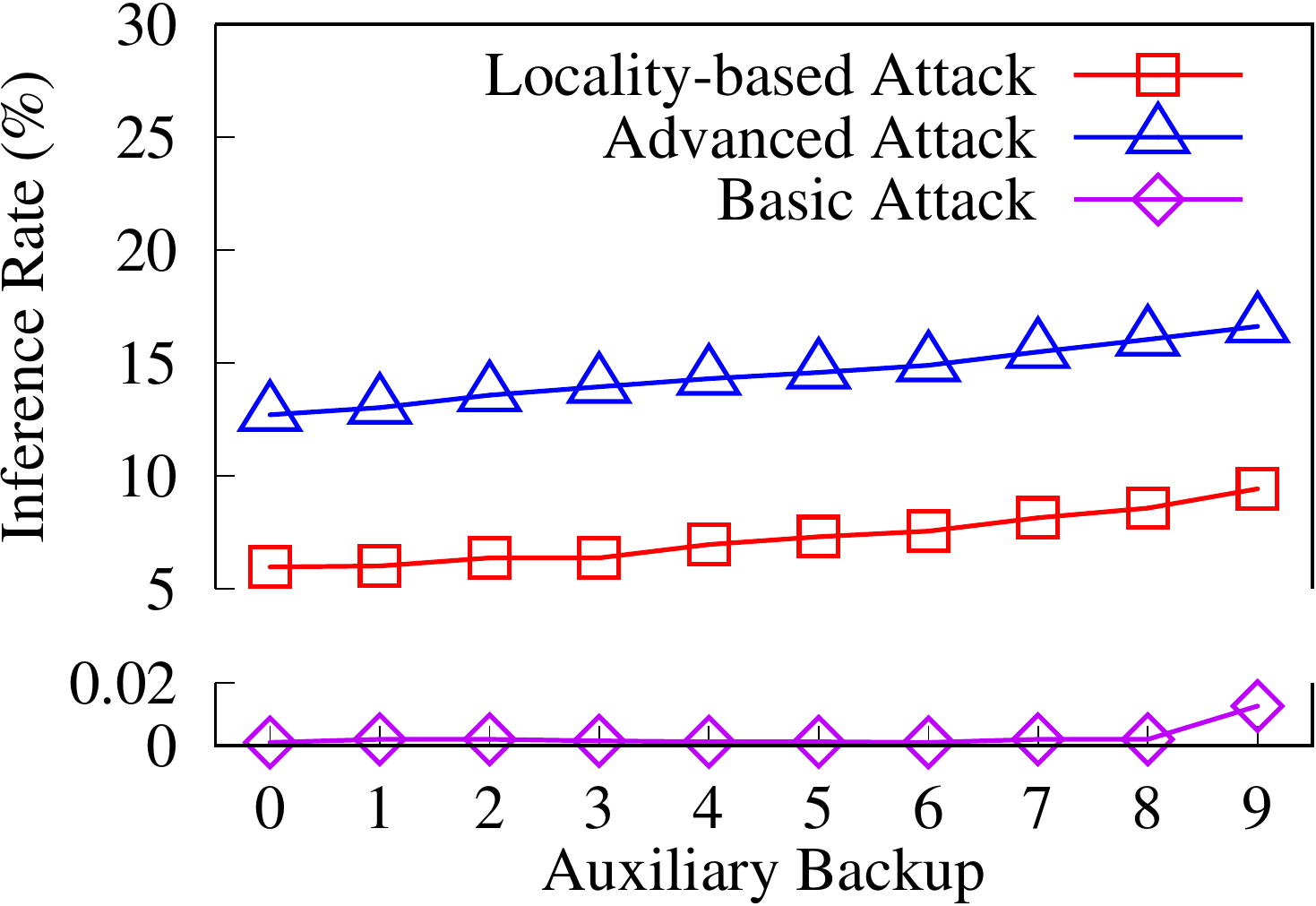}
}
\subfigure[VM dataset]{
\includegraphics[height=1.45in]{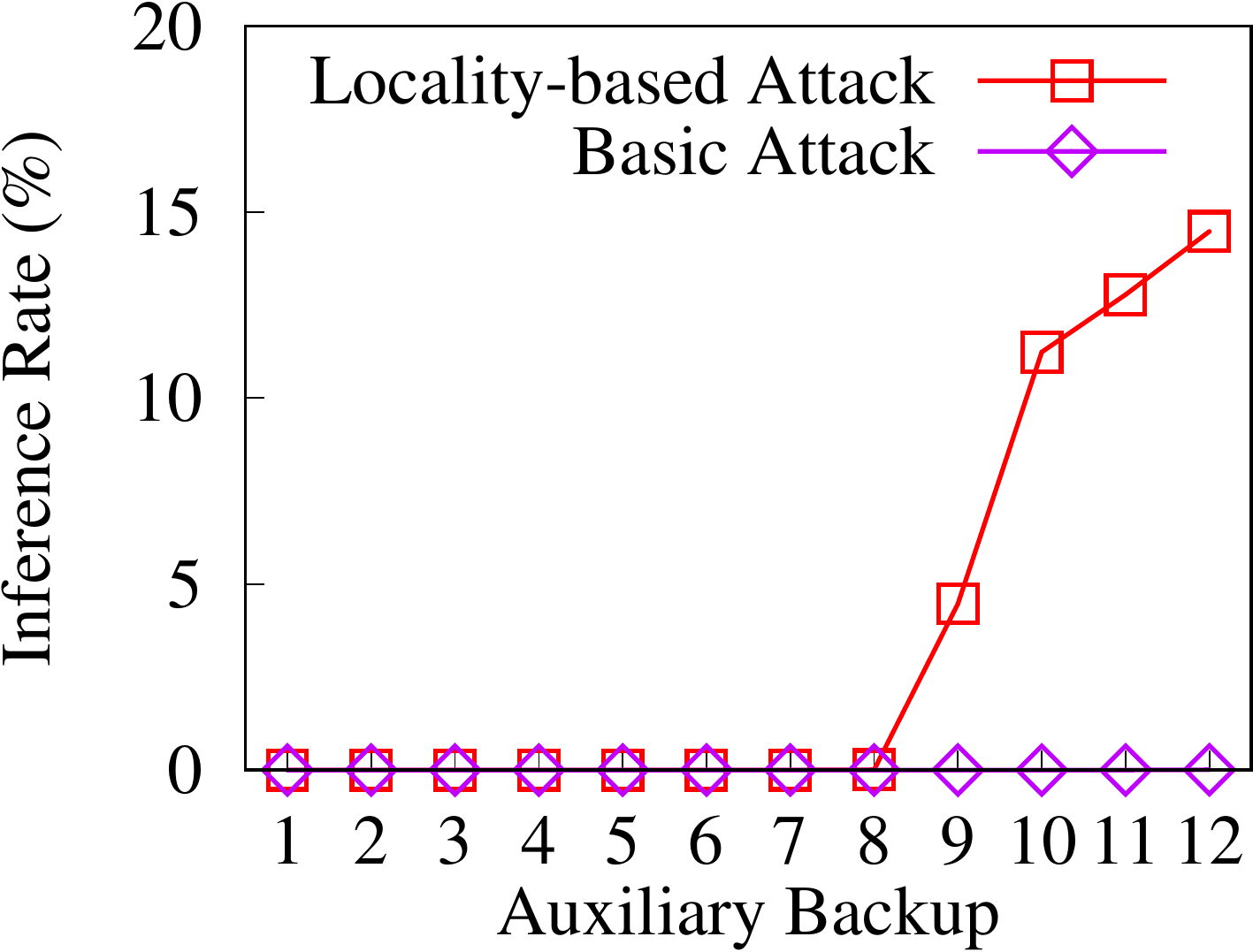}
\label{fig:inference-cipher-only-vm}
}
\vspace{-6pt}
\caption{Attack evaluation: Inference rate in ciphertext-only mode (varying
auxiliary backups, fixed target backup). Note that the locality-based attack
and the advanced locality-based attack are equivalent for the VM dataset.}
\label{fig:infvsaux}
\end{figure*}

Figure~\ref{fig:infvsaux} shows the inference rates of the attacks versus
different auxiliary backups for different datasets (note that the zeroth
auxiliary backup in the synthetic dataset is the publicly available image
snapshot).  As expected, the inference rates of all attacks increase as we use
more recent auxiliary backups, which generally have higher content redundancy
with the target latest backup.  The basic attack is ineffective in all cases,
as the inference rate is no more than 0.0001\% for the FSL dataset, 0.02\% for
the synthetic dataset, and 0.005\% for the VM dataset.  In contrast, both the
locality-based attack and the advanced locality-based attack achieve
significantly high inference rates.  For example, if we use the most recent
prior backup (e.g, the FSL backup on April~21 and the 9th synthetic backup) as
the auxiliary information, the inference rates of the locality-based attack and
the advanced locality-based attack reach as high as 23.2\% and 33.6\% for the
FSL dataset, as well as 9.4\% and 16.6\% for the synthetic dataset,
respectively.  
	
For the VM dataset, the locality-based attack and the advanced locality-based
attack are equivalent, since all chunks have the same size.  We observe that
when we use the first eight VM backups as the auxiliary information, the inference
rate of the locality-based attack is low (less than 0.005\%). The possible
reason is that users have heavy activities during these weeks, such that these
prior backups have low content redundancy with the target latest backup. After
the 8th backup, the inference rate of the locality-based attack increases and
finally achieves 14.5\%.     

\paragraph{Varying target backups:} We next fix the backup on January~22 in
the FSL dataset, the initial Ubuntu image snapshot in the synthetic dataset,
and the first backup in the VM dataset as the auxiliary information, and
infer the original plaintext chunks in each of the following backups (we call
them {\em target backups}).

\begin{figure*}[!t]
\centering
\subfigure[FSL dataset]{
\includegraphics[height=1.45in]{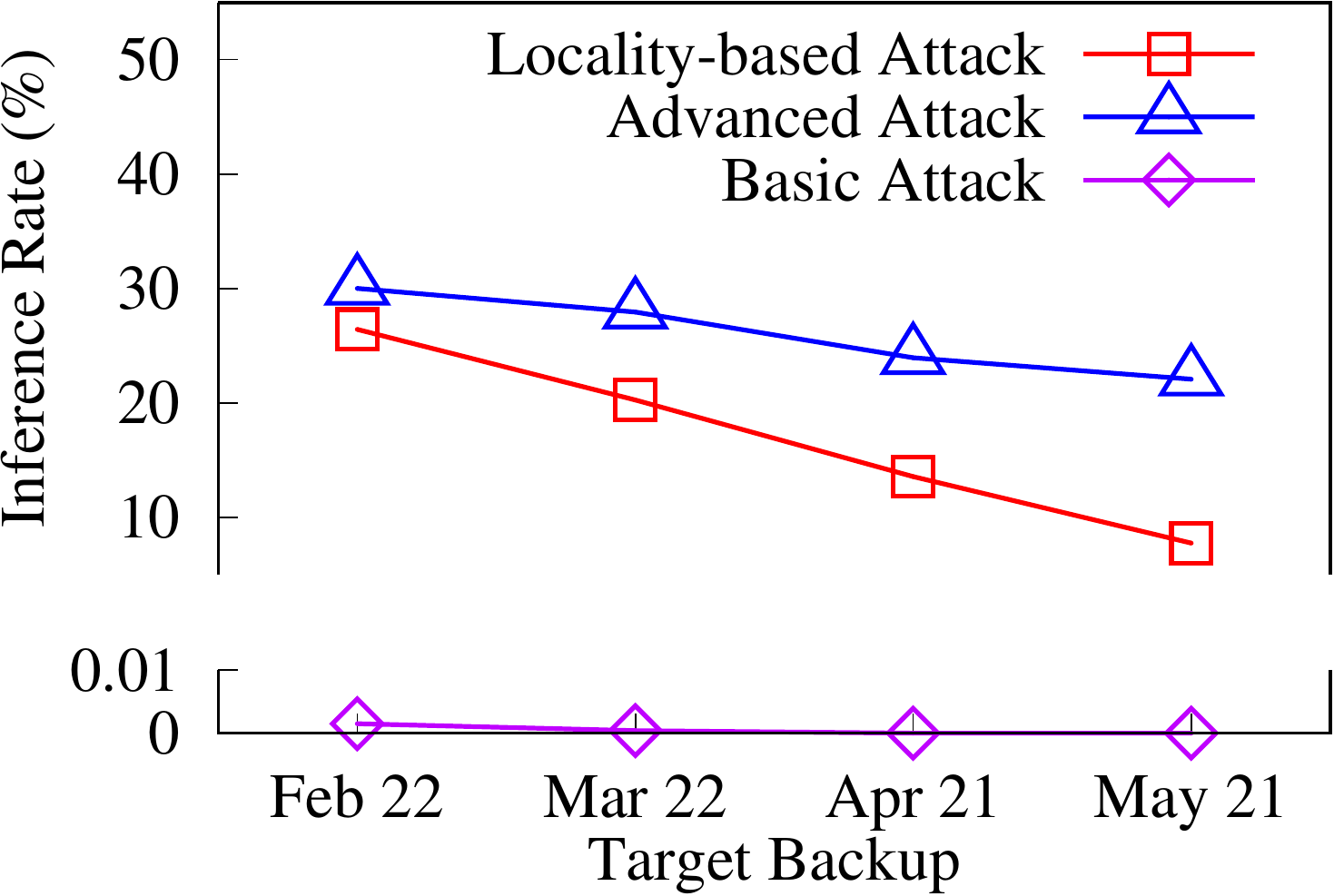}
}
\subfigure[Synthetic dataset]{
\includegraphics[height=1.45in]{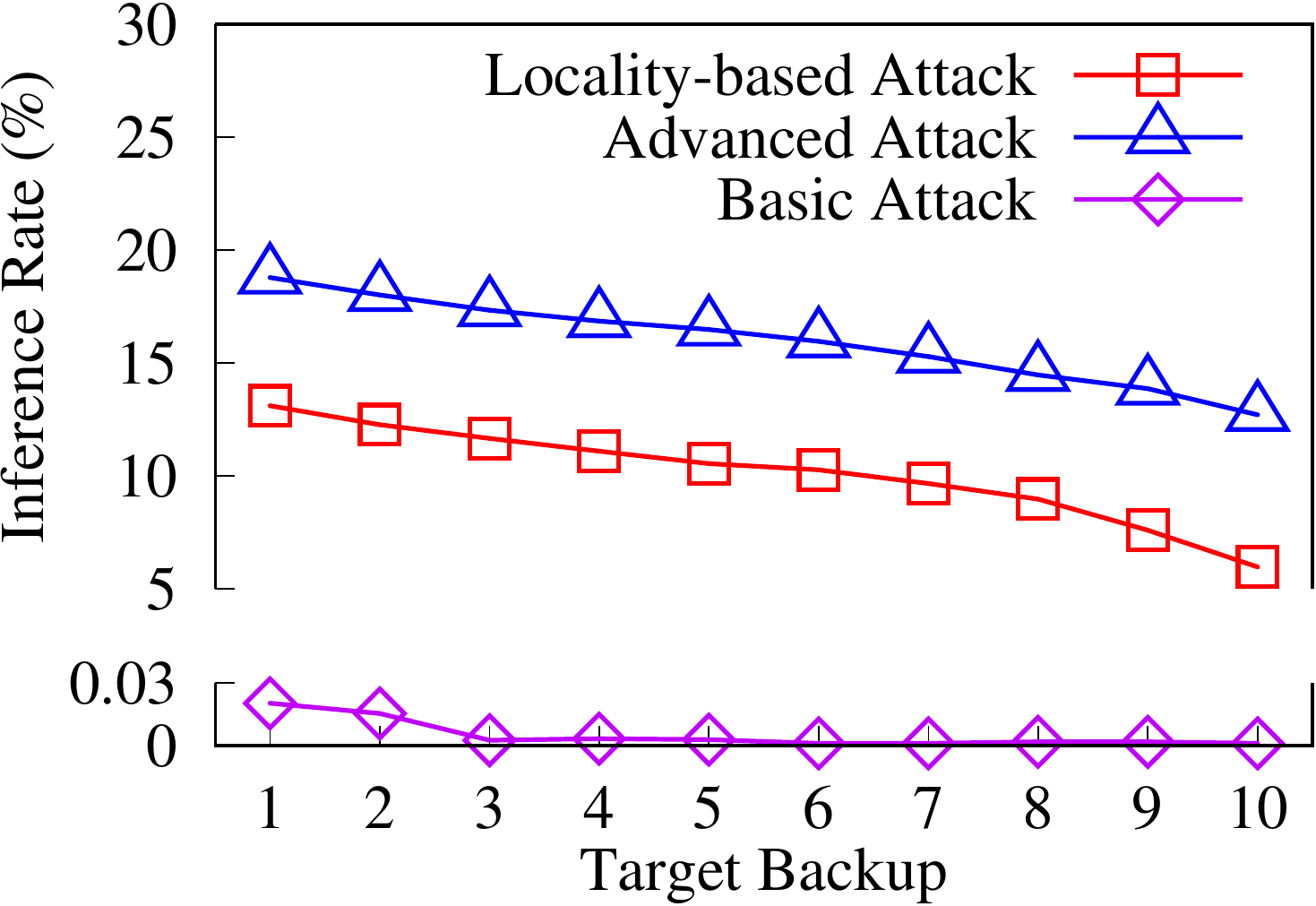}
}
\subfigure[VM dataset]{
\includegraphics[height=1.45in]{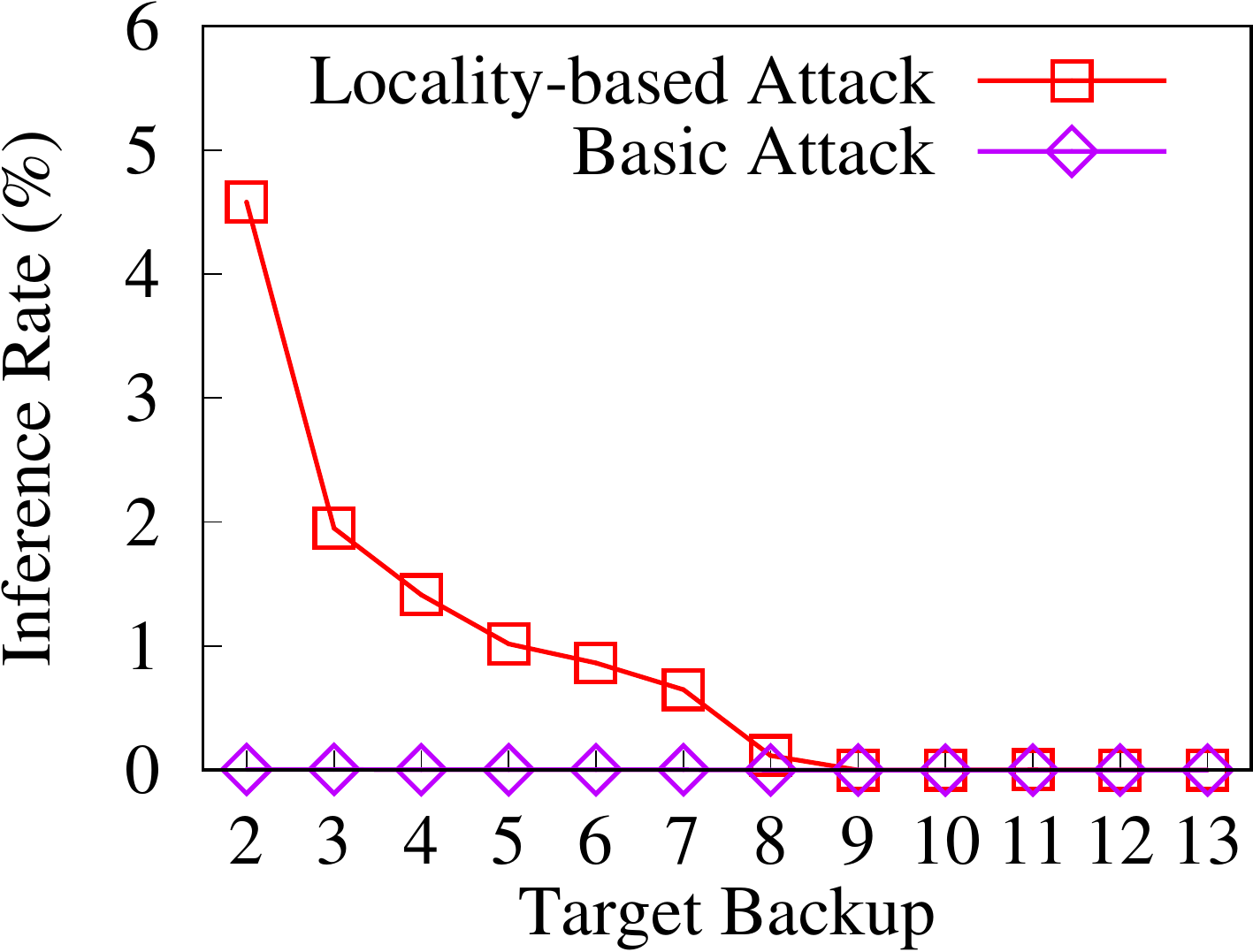}
}
\vspace{-6pt}
\caption{Attack evaluation: Inference rate in ciphertext-only mode (fixed
auxiliary backup, varying target backups).} 
\label{fig:infvstar}
\end{figure*}

Figure~\ref{fig:infvstar} shows the inference rates of the attacks versus
different target backups for different datasets.  Both the locality-based
attack and the advanced locality-based attack are again more severe than the
basic attack, whose inference rate is less than 0.03\%.  For example, the
locality-based attack and the advanced locality-based attack can reach the
inference rates of 26.4\% and 30.0\% for the backup on February~22 in the FSL
dataset, as well as 13.1\% and 18.8\% for the first backup in the synthetic
dataset, respectively.  Since more chunks have been updated after a number of
backups (e.g., four backups for the FSL dataset and ten backups for the
synthetic dataset), the inference rates of the locality-based attack
and the advanced locality-based attack drop to 7.7\% and 22.1\% for the FSL
dataset, as well as 6.0\% and 12.7\% for the synthetic dataset, respectively.  

For the VM dataset, the inference rate of the locality-based attack becomes
low (e.g., around 0.1\%) after using the 8th weekly backup as the target
backup, since heavy updates appear during this period. Nevertheless, the
locality-based attack still achieves a higher inference rate than the basic
attack, whose inference rate is below 0.001\%.      

\paragraph{Attacks over a sliding window:} Finally, we consider the launch of
inference attacks based on a sliding window approach.  Specifically, we choose
the $t$-th backup as the auxiliary information, and infer the original plaintext
chunks in the $(t+s)$-th backup, while we vary $s$ and $t$ in our evaluation.
We mainly focus on the locality-based attack and the advanced locality-based
attack, since the basic attack has low severity.

\begin{figure*}[!t]
\centering
\subfigure[FSL dataset]{
\includegraphics[height=1.45in]{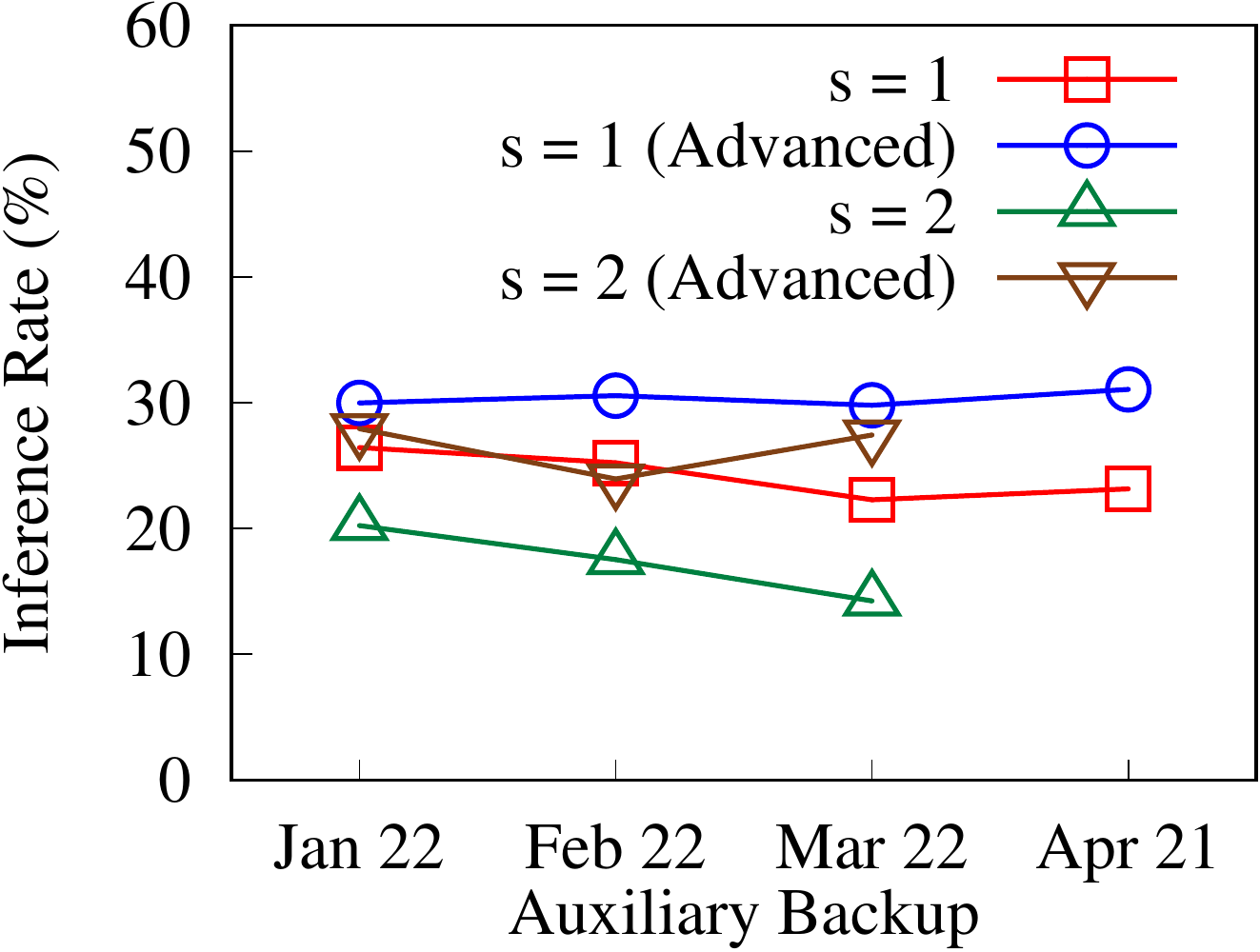}
}
\subfigure[Synthetic dataset]{
\includegraphics[height=1.45in]{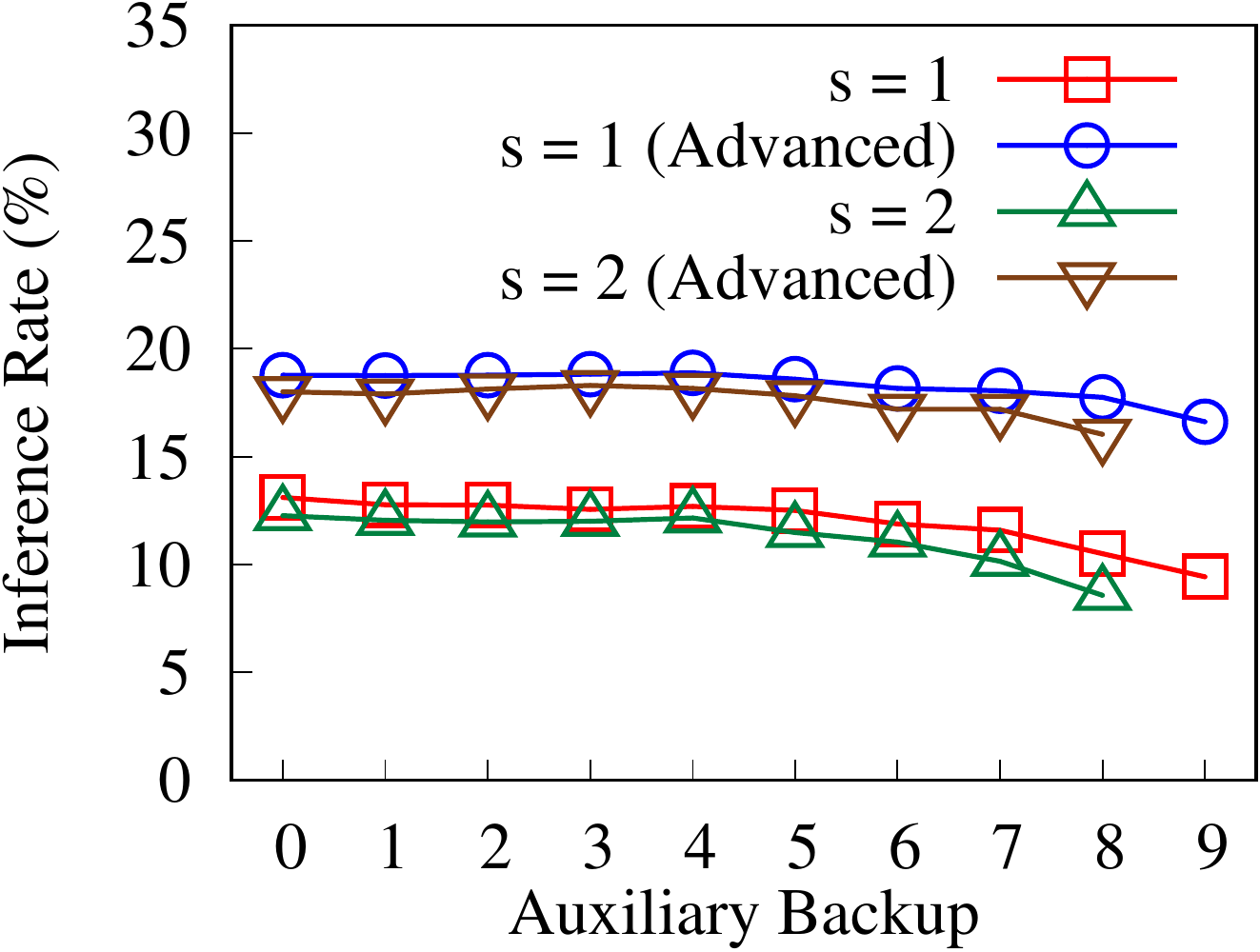}
}
\subfigure[VM dataset]{
\includegraphics[height=1.45in]{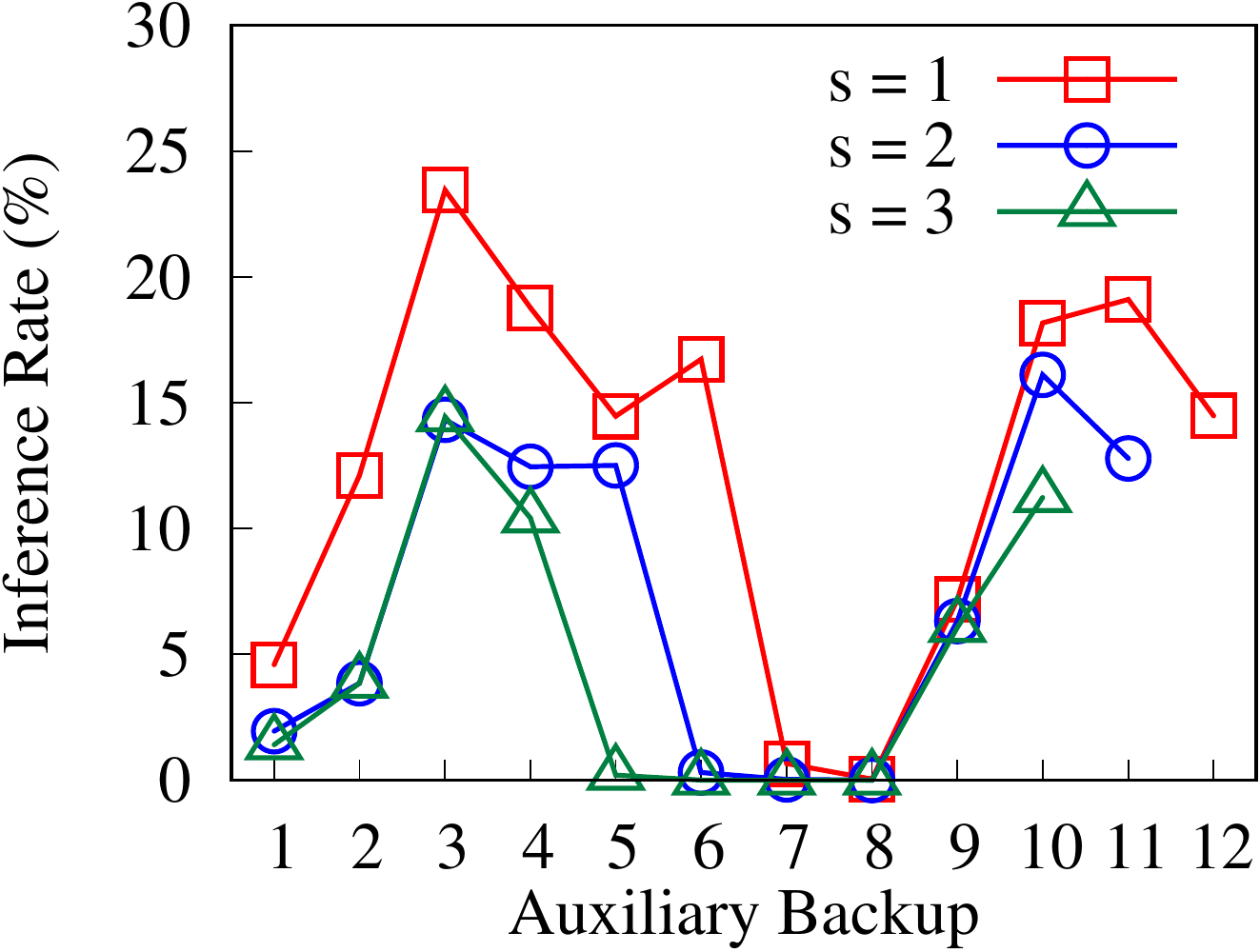}
}
\vspace{-6pt}
\caption{Attack evaluation: Inference rate in ciphertext-only mode 
(over a sliding window).} 
\label{fig:infvsstep}
\end{figure*}

Figure~\ref{fig:infvsstep} shows the inference rates for different $s$, where
the x-axis represents different values of $t$.  The advanced locality-based
attack is more severe than the locality-based attack. For example, in the FSL
dataset, the average inference rates of the locality-based attack are 24.3\%
and 17.3\% for $s=$~1 and $s=$~2, while the corresponding inference rates of
the advanced locality-based attack increase to 30.4\% and 26.4\%,
respectively; in the synthetic dataset, the average inference rates of the
locality-based attack are 12.0\% and 11.3\% for $s=$~1 and $s=$~2, while the
corresponding inference rates of the advanced locality-based attack increase
to 18.3\% and 17.6\%, respectively.

In addition,  the inference rates of the VM dataset fluctuate significantly.
For example, when we use the 3rd weekly backup as the auxiliary information, the
inference rates hit the highest at 23.5\%, 14.3\%, and 14.4\% for $s=$~1, 2,
and 3, respectively. On the other hand, the inference rates drop down to less
than 0.6\% when we use the 5th to 8th weekly backups as the auxiliary information.
Even with such non-preferable cases, the locality-based attack still achieves
moderate severity in general, with an average inference rate of 12.5\%, 7.3\%,
and 4.8\% for $s=$~1, 2, and 3, respectively.

\subsubsection{Inference Rate in Known-Plaintext Mode}

We further evaluate the severity of the locality-based attack and the advanced
locality-based attack in known-plaintext mode.  To quantify the amount of
leakage about the latest backup (see Section~\ref{sec:threat}), we define the
\emph{leakage rate} as the ratio of the number of ciphertext-plaintext chunk
pairs known by the adversary to the total number of ciphertext chunks in a
target backup.  We configure $u=$~1, $v=$~15, and $w=$~500,000.  Note that we
increase $w$ to 500,000 (as opposed to $w=$~200,000 in
Section~\ref{subsubsec:ciphertext}), as the attack in known-plaintext mode can
infer much more ciphertext-plaintext chunk pairs across iterations.  Thus, we
choose a larger $w$ to include them into the inferred set.

\paragraph{Varying leakage rates:}  We fix both the auxiliary and target
backups, and evaluate the inference rates of the attacks for different leakage
rates.  For the FSL dataset, we choose the backup on March~22 as the auxiliary
information to infer the latest backup on May~21; for the synthetic dataset,
we use the initial snapshot as the auxiliary information to infer the 5th
backup snapshot; for the VM dataset, we use the 9th weekly backup as the
auxiliary information to infer the 13th weekly backup.  

\begin{figure}[t]
\centering
\includegraphics[width=3in]{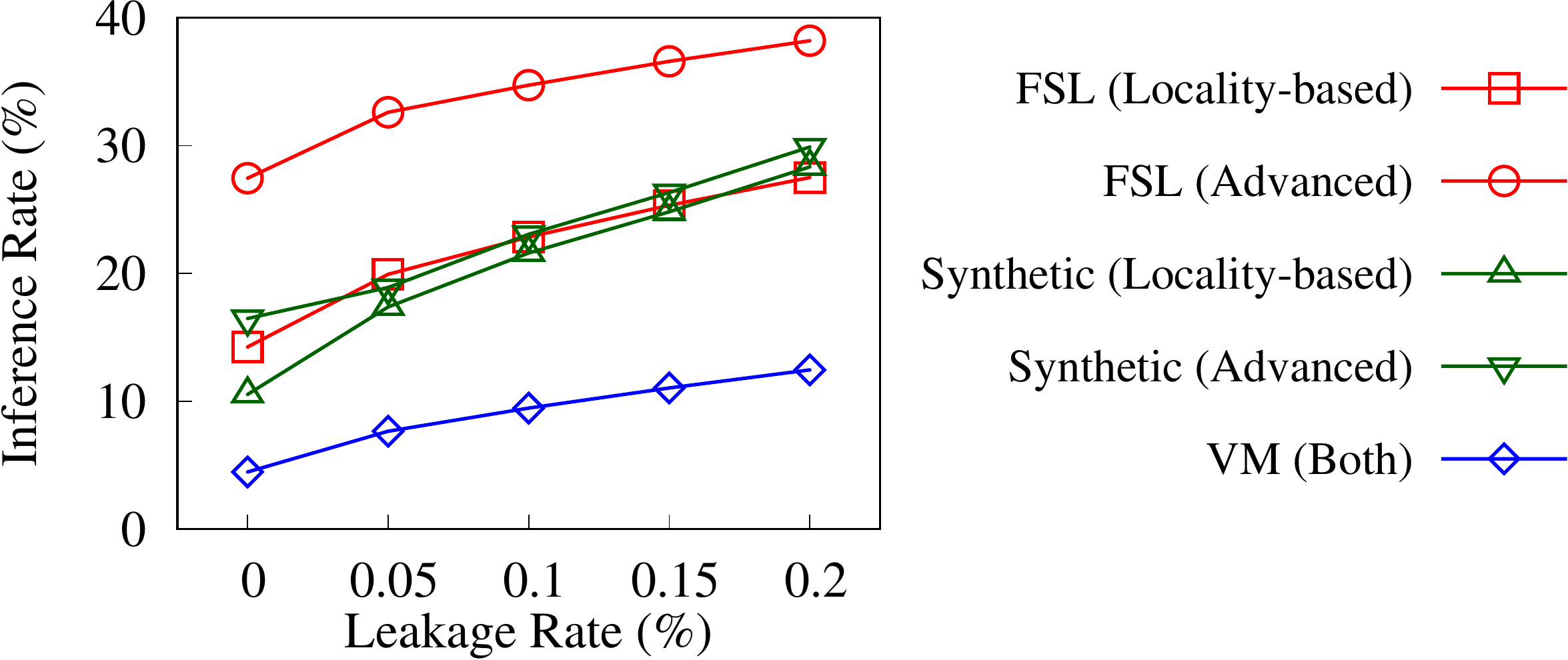}
\vspace{-6pt}
\caption{Attack evaluation: Inference rate in known-plaintext mode (varying
leakage rates).}
\label{fig:revert-with-leak}
\end{figure}

Figure~\ref{fig:revert-with-leak} shows the inference rates of the attacks
(which also include the chunks that are already leaked in known-plaintext
mode) for different leakage rates (varied from 0 to 0.2\%) about the target
backup being inferred. The slight increase in the leakage rate can lead to a
significant increase in the inference rate. For example, when the leakage rate
increases to 0.2\%, the inference rates of the locality-based attack and the
advanced locality-based attack reach 27.5\% and 38.2\% for the FSL dataset,
and 28.3\% and 29.9\% for the synthetic dataset, respectively. Since the VM
dataset includes fixed-size chunks, both attacks incur the same inference rate
of 12.5\% in this case. 

\begin{figure*}[!t]
\centering
\subfigure[FSL dataset]{
\includegraphics[height=1.45in]{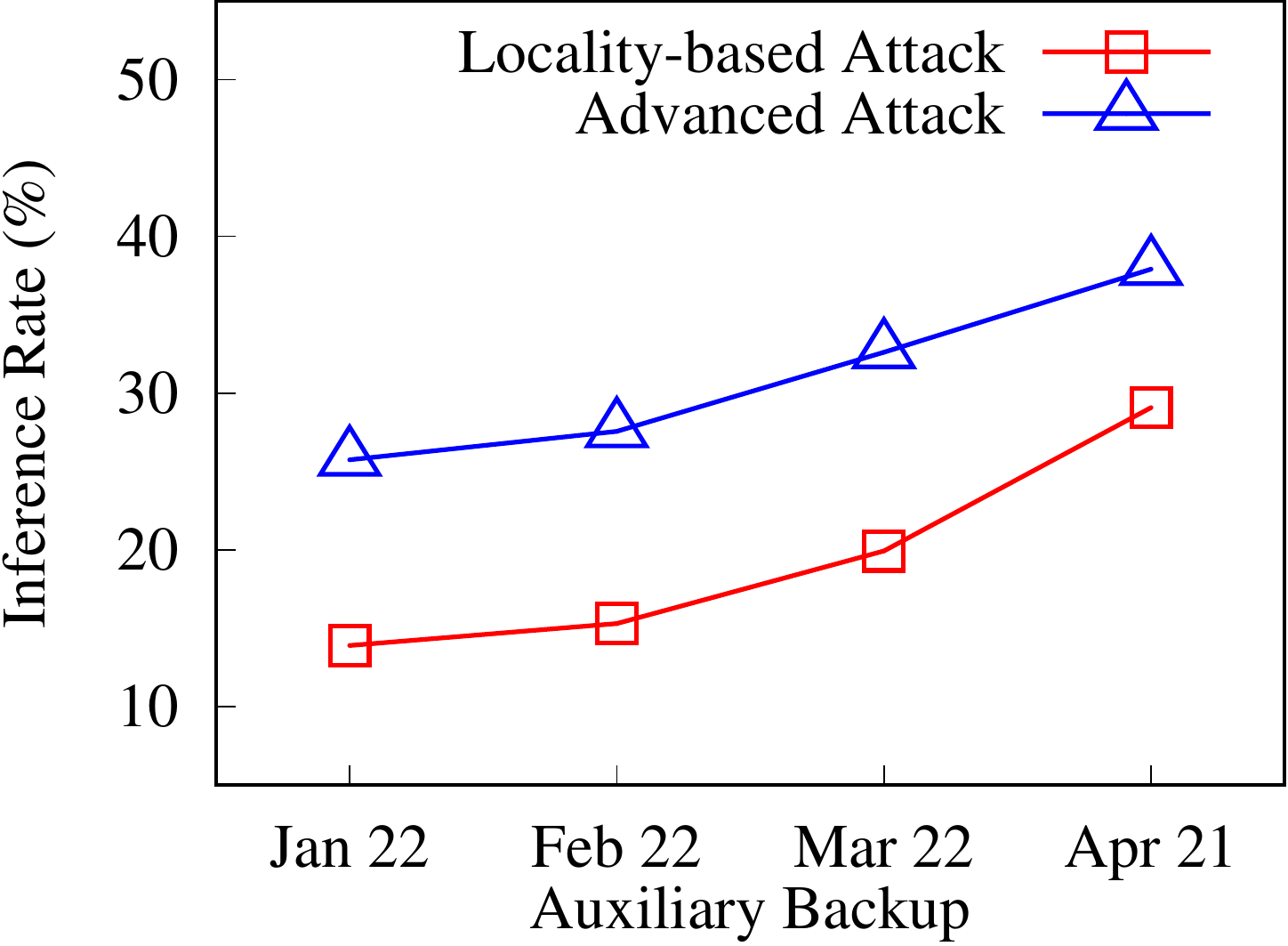}
}
\subfigure[Synthetic dataset]{
\includegraphics[height=1.45in]{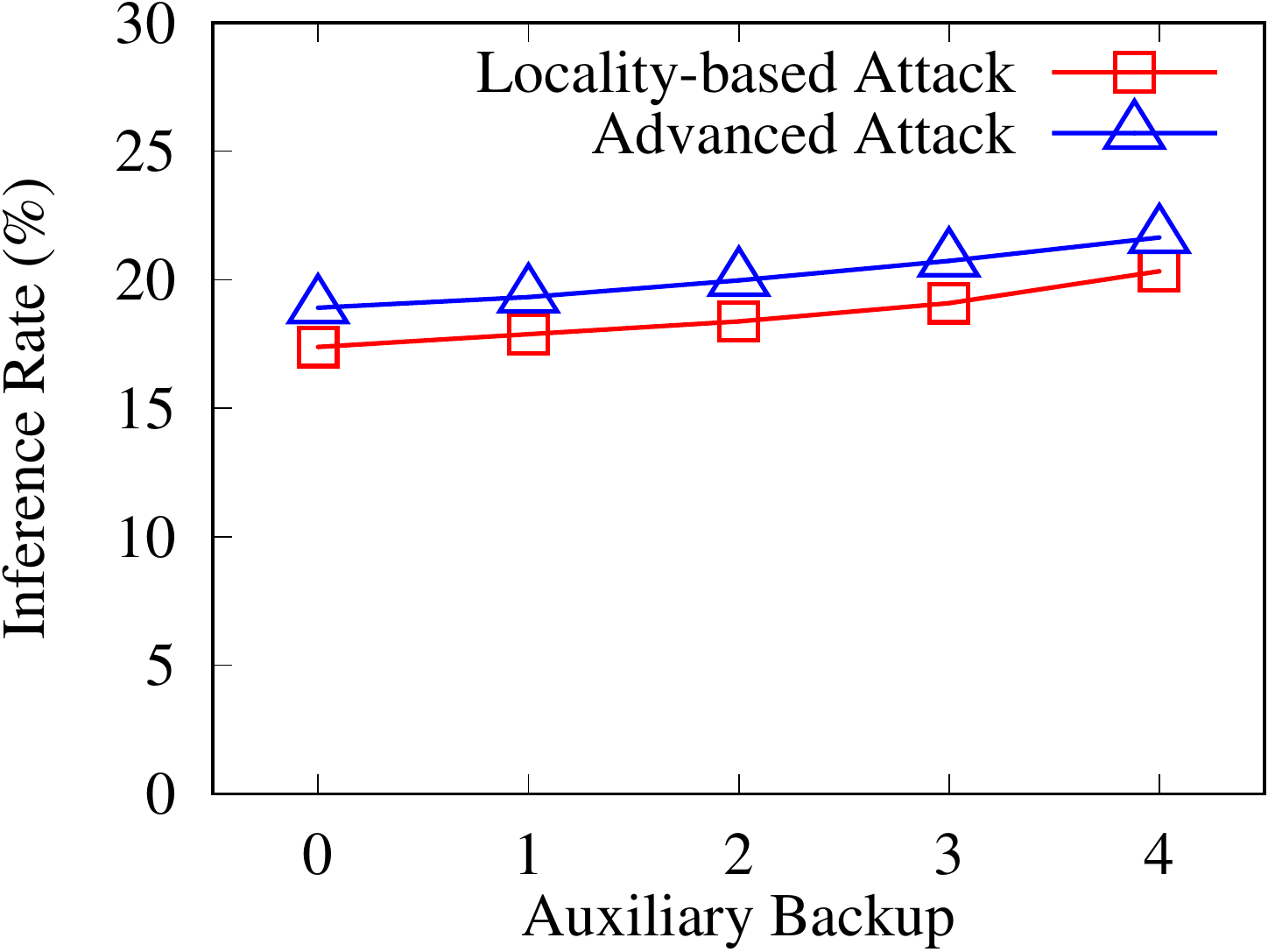}
}
\subfigure[VM dataset]{
\includegraphics[height=1.45in]{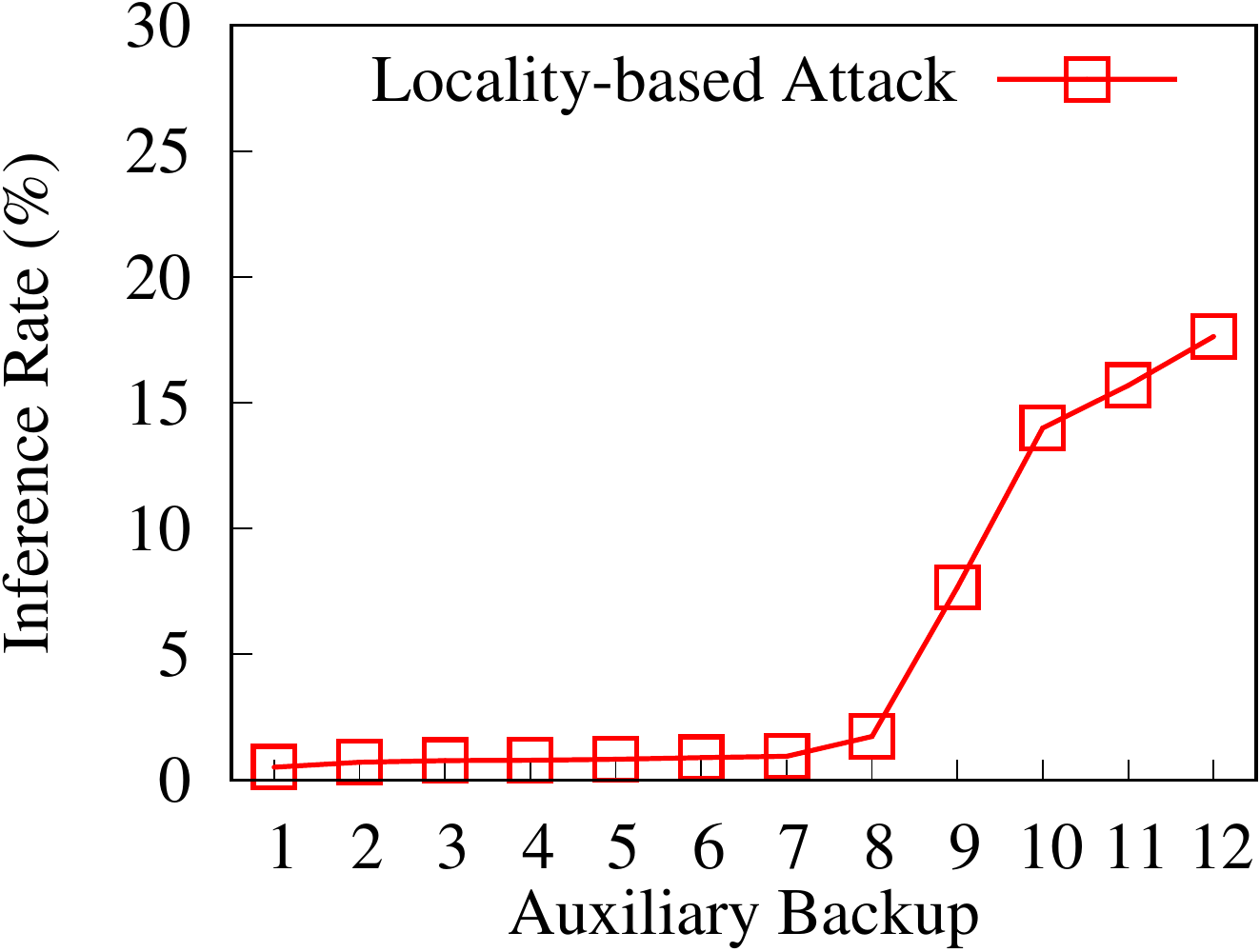}
}
\vspace{-6pt}
\caption{Attack evaluation: Inference rate in known-plaintext mode (varying
auxiliary backups).}
\label{fig:leakage-backup}
\end{figure*}

\paragraph{Varying auxiliary backups:} Based on the above setting, we consider
different prior backups as the auxiliary information, while we fix the target
backup as in the above experiment.  We also fix the leakage rate as 0.05\%. 

Figure~\ref{fig:leakage-backup} shows the inference rates of the attacks for
different auxiliary backups.  We observe a similar tendency as in
Figure~\ref{fig:infvsaux}. Specifically,   in the FSL dataset, when using the
FSL backup on April~21 as the auxiliary information, the inference rates of
the locality-based attack and the advanced locality-based attack achieve
29.1\% and 37.9\%, respectively; in the synthetic dataset, when using the 4th
synthetic backup as the auxiliary information, the corresponding inference
rates achieve 20.3\% and 21.6\%, respectively; in the VM dataset, when using
the 12th VM backup as the auxiliary information, the inference rate achieves
17.6\% (same in both the locality-based attack and the advanced locality-based
attack). 

\section{Defenses}
\label{sec:defense}

The deterministic nature of encrypted deduplication discloses the frequency
distribution of the underlying plaintext chunks, thereby making frequency
analysis feasible.  To defend against frequency analysis, we consider two
defense approaches, namely {\em MinHash encryption} and {\em scrambling}. 

\subsection{MinHash Encryption}

MinHash encryption builds on Broder's theorem \cite{broder97}, which states
that if two sets share a large fraction of common elements (i.e., they are
highly similar), then the probability that both sets share the same minimum
hash element is also high.  Since two backups from the same data source are
expected to be highly similar and share a large number of identical chunks
\cite{wallace12}, MinHash encryption leverages this property to perform
encrypted deduplication in a different way from the original MLE
\cite{bellare13a,bellare13b}.  We emphasize that previous deduplication
approaches also leverage Broder's theorem to minimize the memory usage of the
fingerprint index in plaintext deduplication \cite{bhagwat09,xia11} or key
generation overhead in server-aided MLE \cite{qin17}.  Also, security analysis
shows that MinHash encryption preserves data confidentiality as in
server-aided MLE \cite{qin17}.  Thus, we do not claim the novelty of the
design of MinHash encryption.  Instead, our contribution is to study its
effectiveness in defending against frequency analysis. 

Algorithm~\ref{alg:tun-enc} shows the pseudo-code of MinHash encryption, while
we elaborate the implementation details in Section~\ref{subsec:def_impl}.  
MinHash encryption takes a sequence of plaintext chunks $\mathbf{M}$ as input,
and returns a sequence of ciphertext chunks $\mathbf{C}$ as output.  It
partitions the plaintext chunks into {\em segments} (Line~3), 
each of which is a non-overlapped sub-sequence of adjacent plaintext chunks.
For each segment $S$, MinHash encryption computes the minimum fingerprint $h$
of all chunks in $S$, using the fingerprint value as the hash value of each
chunk.  It derives the segment-based key $K_S$ based on $h$ (Line~6), for
example, by querying the key manager as in DupLESS \cite{bellare13b}
(see Section~\ref{subsec:encrypted-dedup}).  It then encrypts each chunk in
$S$ using $K_S$ and adds the resulting ciphertext chunk to $\mathbf{C}$
(Lines~7-10).  Finally, it returns $\mathbf{C}$ (Line~12).  Note that MinHash
encryption only requests keys on a per-segment basis rather than on a
per-chunk basis. As the number of segments is much less than that of chunks,
the key generation overhead is greatly mitigated \cite{qin17}.  

\begin{algorithm}[t]
\caption{MinHash Encryption} 
\label{alg:tun-enc}
\small
\begin{algorithmic}[1]
\Procedure{MinHash Encryption}{$\mathbf{M}$}		
  \State Initialize $\mathbf{C}$
  \State Partition $\mathbf{M}$ into segments 
  \For{each segment $S$} 
    \State $h \gets $ minimum fingerprint of all chunks in $S$
    \State $K_S \gets $ segment-based key derived from $h$
	\For{each chunk $M \in S$} 
	   \State $C \gets${\sc Encrypt}$(K_S, M)$ 
	   \State Add $C$ into $\mathbf{C}$
	\EndFor
\EndFor
\State \Return $\mathbf{C}$
\EndProcedure 
\end{algorithmic}
\end{algorithm}

MinHash encryption is robust against the locality-based attack, by (slightly)
breaking the deterministic nature of encrypted deduplication.  Its rationale
is that segments are highly similar as they share many identical plaintext
chunks in backups \cite{bhagwat09,xia11}.  Thus, their minimum fingerprints,
and hence the secret keys derived for segments, are likely to be the same as
well due to Broder's theorem \cite{broder97}.  This implies that most
identical plaintext chunks across segments are still encrypted by the same
secret keys into identical ciphertext chunks, thereby preserving deduplication
effectiveness.  However, some identical plaintext chunks may still reside in
different segments with different minimum fingerprints and hence different
secret keys, so their resulting ciphertext chunks will be different and cannot
be deduplicated, leading to a slight degradation of storage efficiency.
Nevertheless, such ``approximate'' deduplication sufficiently alters the
overall frequency ranking of ciphertext chunks by encrypting a small fraction
of duplicate chunks using different keys, thereby making frequency analysis
ineffective.   

\subsection{Scrambling}
\label{subsec:scrambling}

Scrambling augments MinHash encryption by disturbing the processing sequence
of chunks, so as to prevent an adversary from correctly identifying the
neighbors of each chunk in the locality-based attack.  It is applied before
the chunks are encrypted and stored, and its idea is to scramble the original
plaintext chunk sequence $\mathbf{M}$ into a new sequence $\mathbf{M}'$.  To
be compatible with MinHash encryption, scrambling works on a per-segment basis
by shuffling the ordering of chunks within each segment.  Each plaintext chunk
is still encrypted via MinHash encryption and stored as a ciphertext chunk,
while the original file can still be reconstructed based on its file recipe
and key recipe (see Section~\ref{sec:basics}).  Specifically, the file recipe
contains a list of fingerprints that are stored in the original order of the
plaintext chunks before scrambling.  If a client wants to restore the original
file, it first retrieves the file recipe and the key recipe (which are
encrypted by the client's own secret key), followed by decrypting the
ciphertext chunks (based on the key recipe) and restoring the original order
of the plaintext chunks and hence the original file (based on the file
recipe). 

Note that scrambling does not change the storage efficiency of MinHash
encryption, since it only changes the order of plaintext chunks.  Also, we
apply scrambling on a per-segment basis, while a deduplicated storage system
typically organizes unique chunks in {\em containers} \cite{lillibridge13}
that serve as the basic read/write units.  In our prototype (see
Section~\ref{sec:case}), we configure the container size larger than the
segment size (e.g., we set the container size as 4\,MB, while setting the
maximum segment size as 2\,MB).  Thus, the scrambling approach has limited
impact on the chunk layout across containers, so it does not add substantial
overhead to the overall read/write performance.  In Section~\ref{sec:case}, we
will study how the scrambling approach affects the performance of a
deduplicated storage system.

Algorithm~\ref{alg:tun-scrambling} elaborates the pseudo-code of scrambling.
It first partitions the original plaintext chunk sequence $\mathbf{M}$ into
segments as in MinHash encryption (Line~2).  Then for each chunk of a segment
$S$, the algorithm randomly adds the chunk to either the front of $S'$ or the
end of $S'$, where $S'$ is the scrambled version of $S$ (Lines~6-13).
Finally, it returns the scrambled sequence $\mathbf{M}'$ that includes all the
scrambled segments (Line~16). 
 
\begin{algorithm}[t]
\caption{Scrambling} 
\label{alg:tun-scrambling}
\begin{small}
\begin{algorithmic}[1]
\Procedure{Scrambling}{$\mathbf{M}$}		
	\State Initialize $\mathbf{M}'$
	\State Partition $\mathbf{M}$ into segments 
	\For{each segment $S$} 
	  \State Initialize $S'$
	  \For{each chunk $M \in S$} 
		\State Generate a random number $r$ 
		\If{$r$ is odd}
			\State Add $M$ as first chunk of $S'$
		\Else
			\State Add $M$ as last chunk of $S'$
		\EndIf
	  \EndFor
	   \State Add $S'$ into $\mathbf{M}'$
	\EndFor
	\State \Return $\mathbf{M}'$
\EndProcedure
\end{algorithmic}
\end{small}
\end{algorithm}

\section{Defense Evaluation}
\label{sec:defense_eval}

We conduct trace-driven evaluation on MinHash encryption and scrambling in
three aspects: defense effectiveness, storage efficiency, and metadata access
overhead.  

\subsection{Methodology}
\label{subsec:def_impl}

Since both FSL and VM datasets do not contain actual contents, we simulate
our defense approaches by directly operating on chunk fingerprints.  First, we
identify segment boundaries based on chunk fingerprints, by following the
variable-size segmentation scheme in \cite{lillibridge09}.  Specifically, the
segmentation scheme is configured by the minimum, average, and maximum segment
sizes.  It places a segment boundary at the end of a chunk fingerprint if (i)
the size of each segment is at least the minimum segment size, and (ii) the
chunk fingerprint modulo a pre-defined divisor (which determines the average
segment size) is equal to some constant (e.g., $-1$), or the inclusion of the
chunk makes the segment size larger than the maximum segment size.  In our
evaluation, we set the minimum, average, and maximum segment sizes as 512\,KB,
1\,MB, and 2\,MB, respectively. 

After scrambling the orders of chunks (a.k.a. fingerprints) in each segment,
we mimic MinHash encryption as follows.  We first calculate the minimum chunk
fingerprint $h$ of each segment.  We then concatenate $h$ with each chunk
fingerprint in the segment and compute the SHA-256 hash of the concatenation.
We also truncate the hash result to be consistent with the fingerprint sizes
in the original FSL and VM datasets, respectively.  The truncated hash result
can be viewed as the fingerprint of the ciphertext chunk.  We can easily check
that identical plaintext chunks under the same $h$ will lead to identical
ciphertext chunks that can be deduplicated. 

\subsection{Defense Effectiveness}
\label{subsec:def_eval}

We evaluate our defense schemes, including (i) MinHash encryption only and
(ii) the combined MinHash encryption and scrambling scheme, against the
advanced locality-based attack in known-plaintext mode under the same
parameter setting as in Section~\ref{sec:attack_result}. Note that the
advanced locality-based attack reduces to the locality-based attack in the VM
dataset, which uses fixed-size chunking. 

\begin{figure*}[t]
\centering
\subfigure[FSL dataset]{
\includegraphics[width=.31\textwidth]{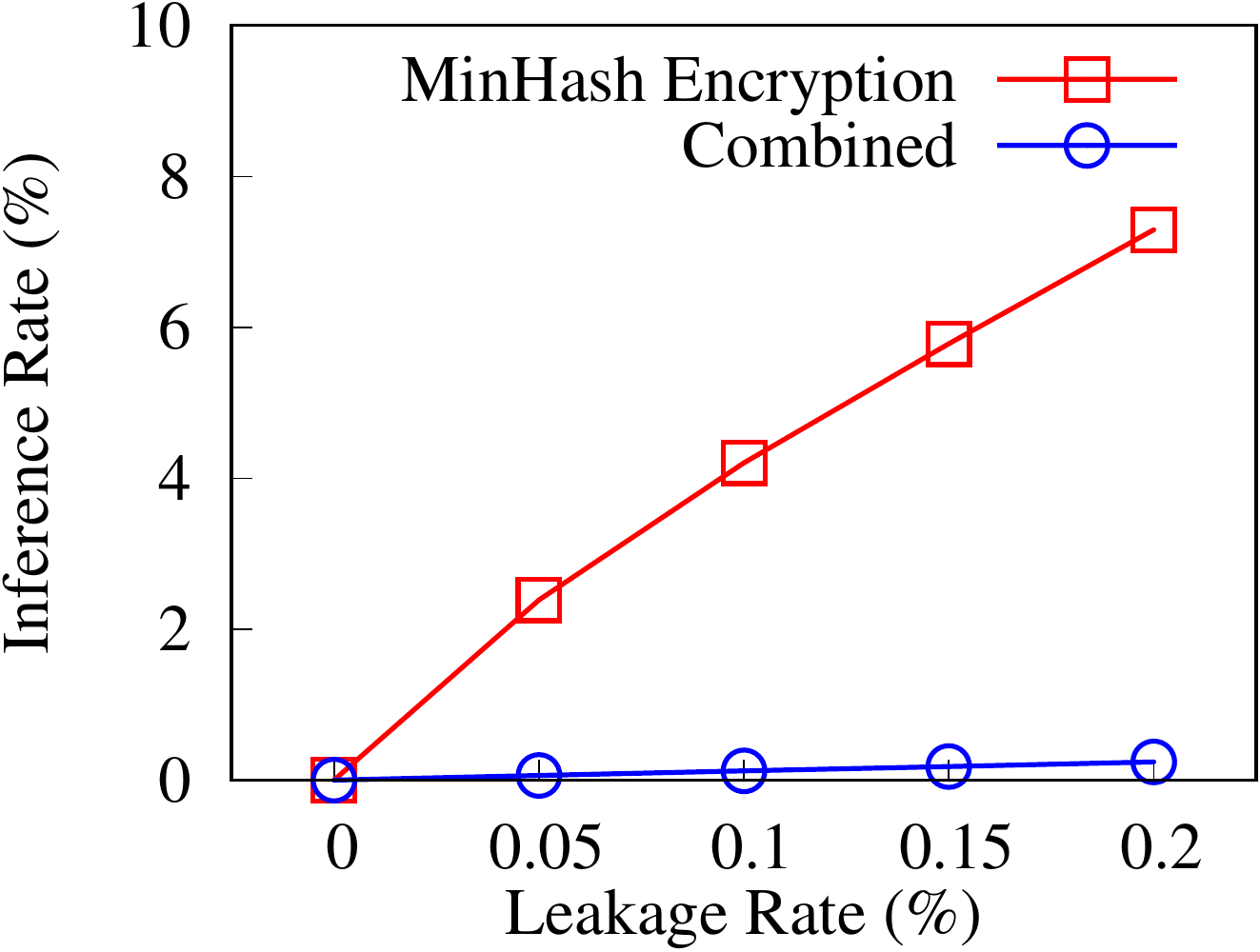}
\label{fig:defense_fsl}
}
\subfigure[Synthetic dataset]{
\includegraphics[width=.31\textwidth]{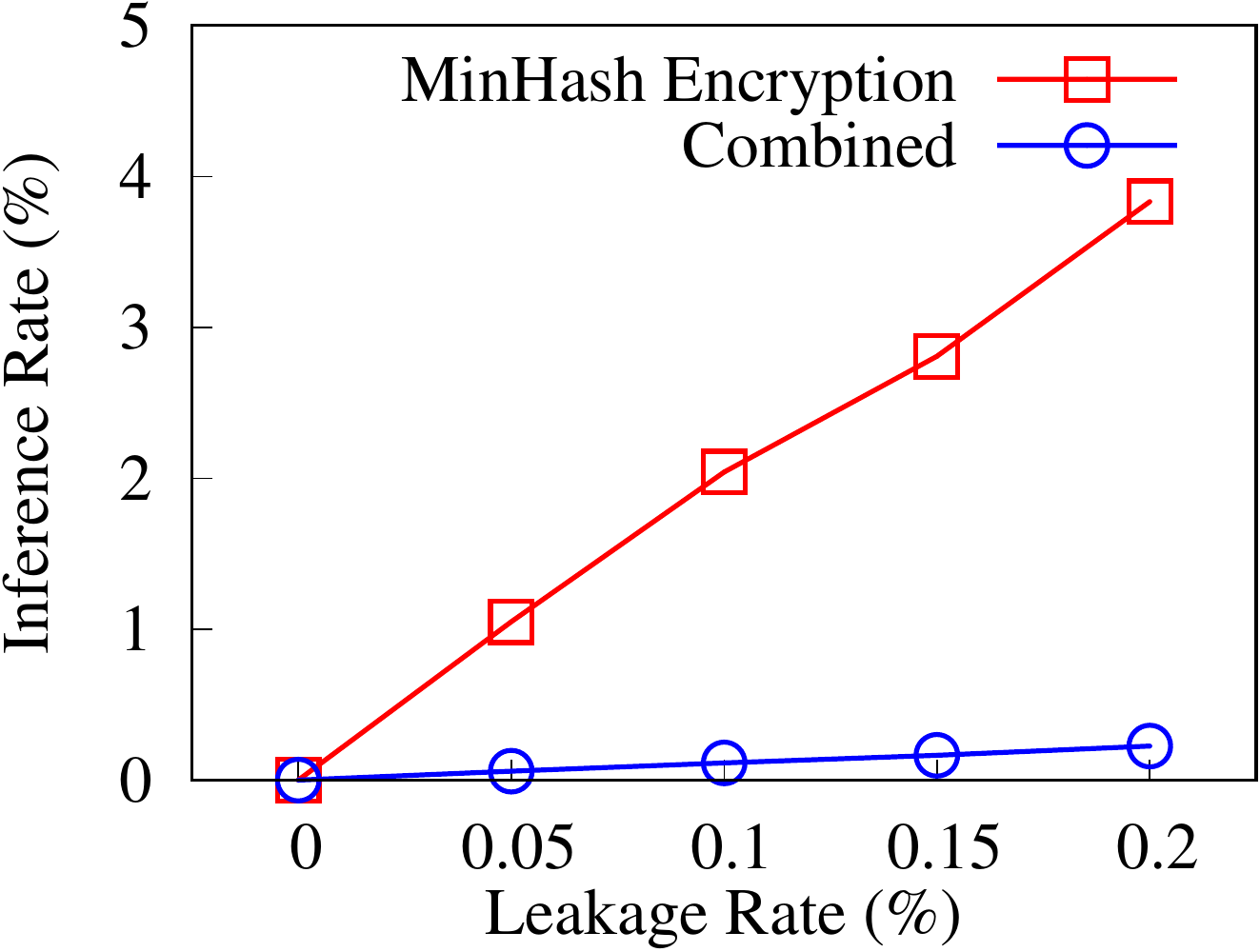}
\label{fig:defense_syn}
}
\subfigure[VM dataset]{
\includegraphics[width=.31\textwidth]{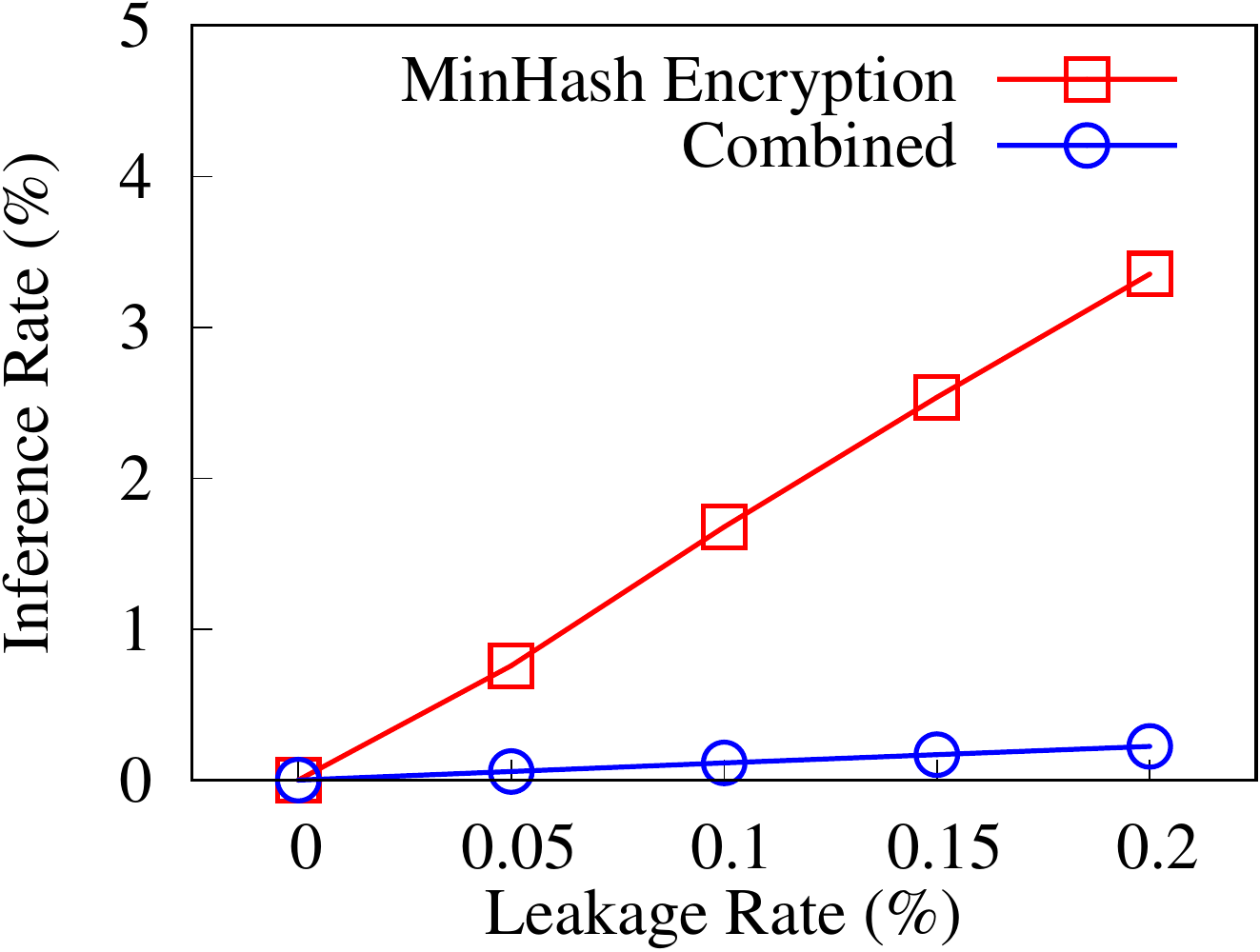}
\label{fig:defense_vm}
}
\vspace{-6pt}
\caption{Defense effectiveness: Inference rate in known-plaintext mode under
MinHash encryption only and the combined MinHash encryption and scrambling
scheme.}
\label{fig:defense}
\end{figure*}

Figure~\ref{fig:defense} shows the inference rate versus the leakage rate.
When the leakage rate is 0.2\%, MinHash encryption suppresses the inference
rate to 7.3\%, 3.8\%, and 3.4\% for the FSL, synthetic, and VM datasets,
respectively, under the advanced locality-based attack.  In addition, the
combined MinHash encryption and scrambling scheme further suppresses the
inference rate to 0.2-0.24\% only for all datasets.  This shows that
scrambling effectively enhances the protection of MinHash encryption, and
the combined scheme effectively defends against the advanced locality-based
attack. 

\subsection{Storage Efficiency}

We evaluate the storage efficiency of the combined MinHash encryption and
scrambling scheme.  Specifically, we add the encrypted backups to storage in
the order of their creation times, and measure the storage saving as the
percentage of the total size of all ciphertext chunks reduced by
deduplication.  We compare the storage saving with that of the original MLE,
which performs chunk-based deduplication that operates at the more
fine-grained chunk level and eliminates all duplicate chunks. Here, we do not
consider the metadata overhead.

\begin{figure*}[!t]
\centering
\subfigure[FSL dataset]{
\includegraphics[width=.31\textwidth]{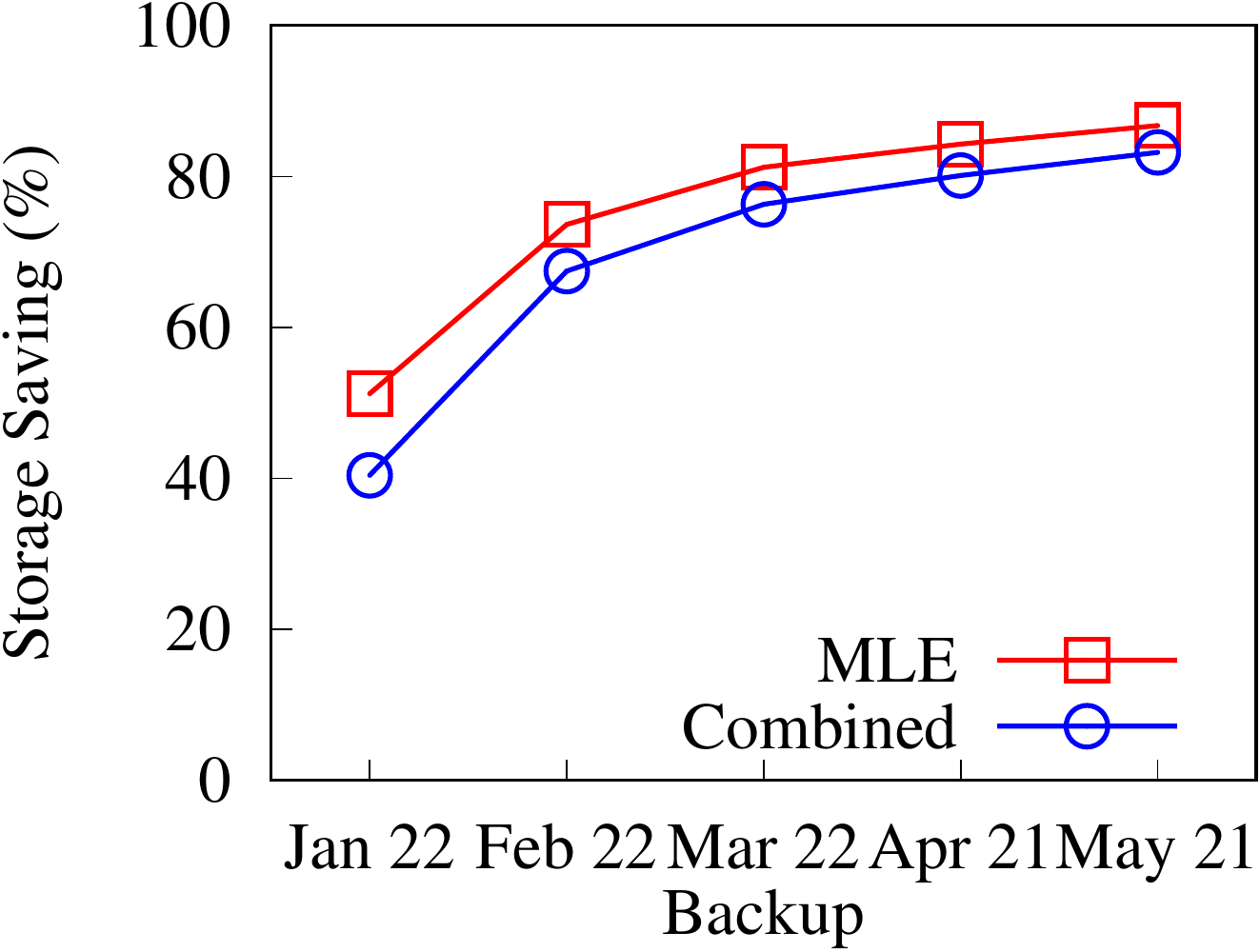}
\label{fig:storage_fsl}
}
\subfigure[Synthetic dataset]{
\includegraphics[width=.31\textwidth]{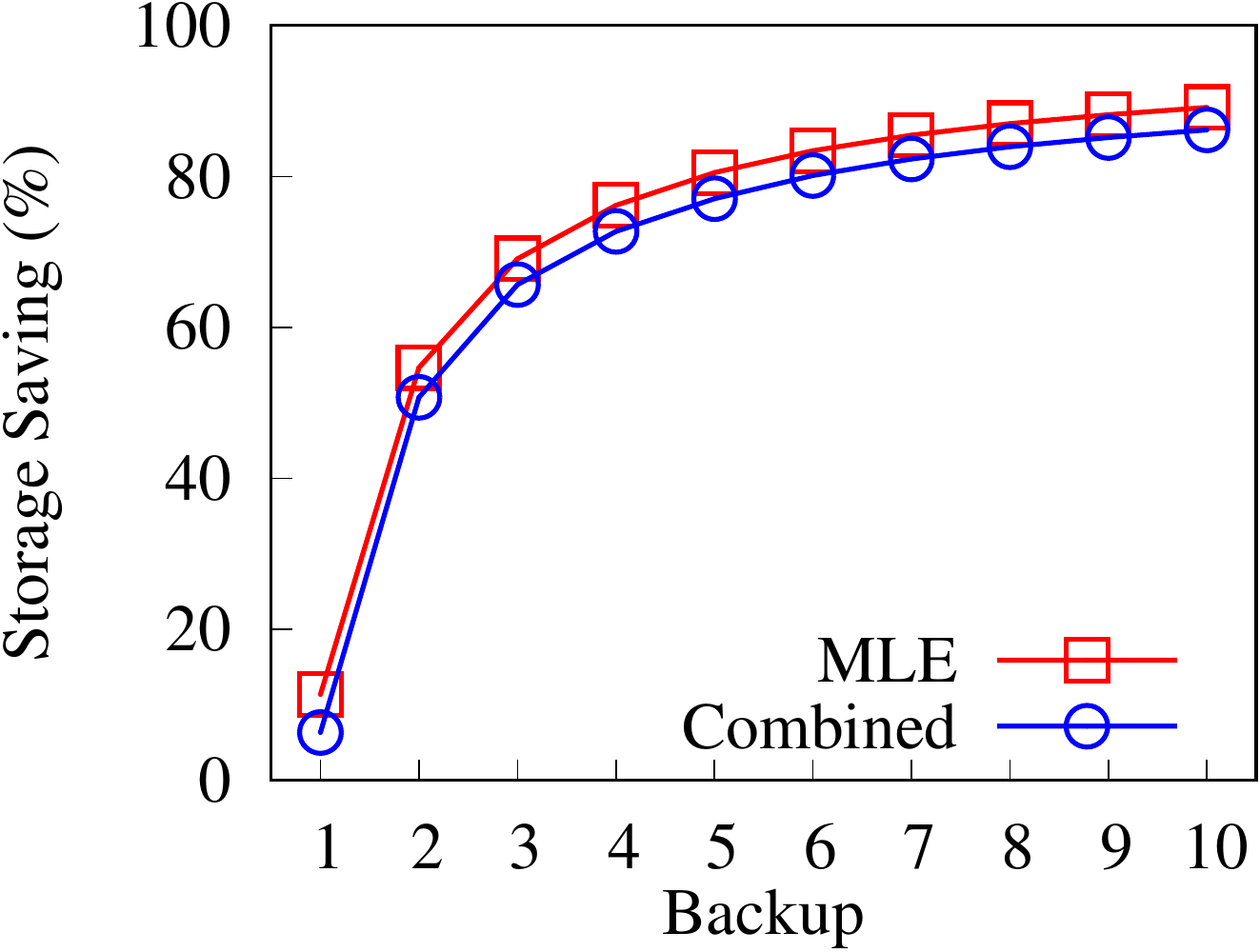}
\label{fig:storage_syn}
}
\subfigure[VM dataset]{
\includegraphics[width=.31\textwidth]{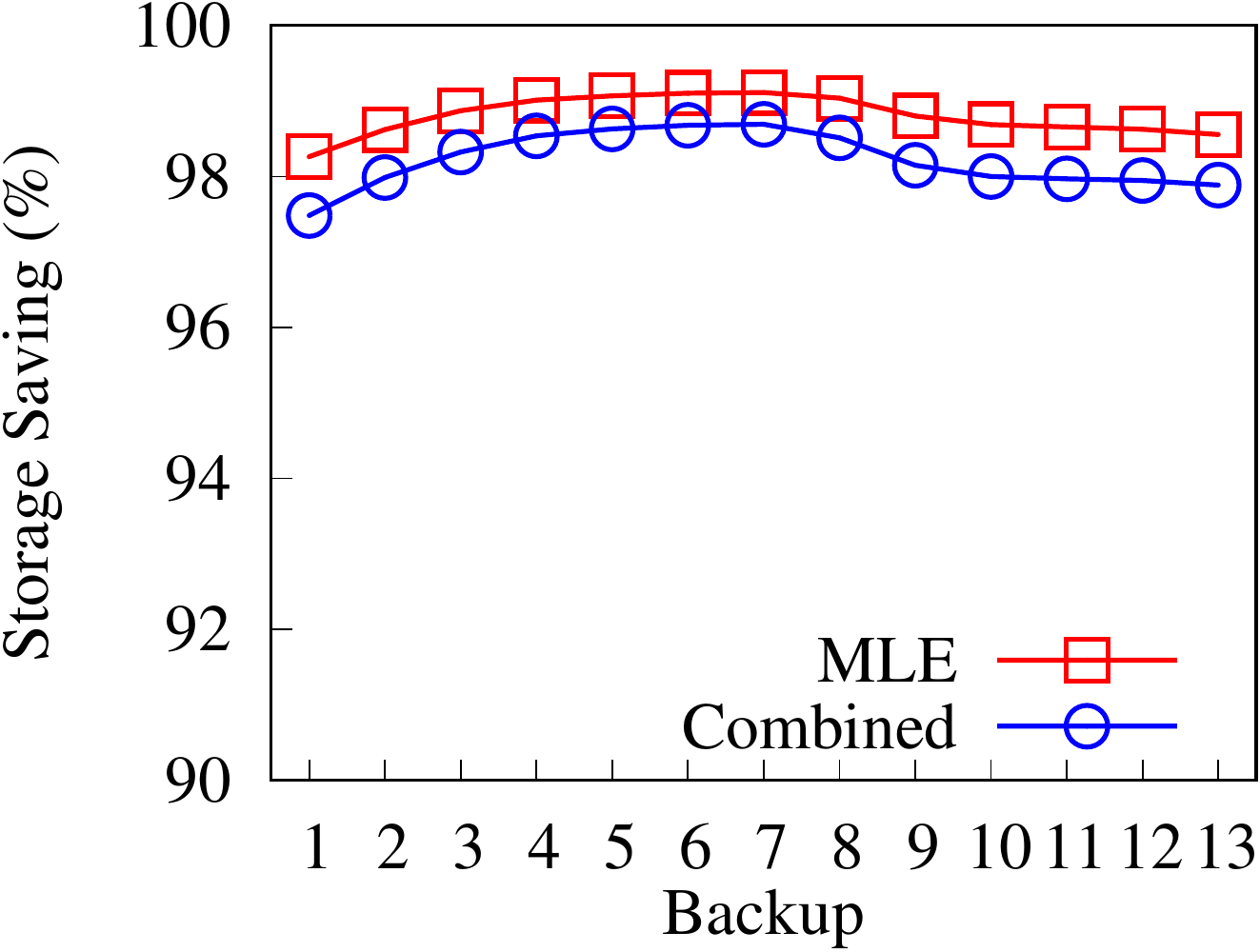}
\label{fig:storage_vm}
}
\vspace{-6pt}
\caption{Storage efficiency of the combined MinHash encryption and scrambling
scheme.}
\end{figure*}

Figure~\ref{fig:storage_fsl} shows the storage saving after storing each FSL
backup. After storing all five backups, the combined scheme achieves a storage
saving of 83.2\% (which corresponds to a deduplication ratio of
6.0$\times$), which is 3.6\% less than that of MLE. 

Figure~\ref{fig:storage_syn} shows the storage saving after storing each
synthetic snapshot. After 11 backups, the combined scheme achieves a storage
saving of 86.2\% (which corresponds to the deduplication ratio of
7.2$\times$). The drop of the storage saving is about 3\% compared to MLE,
which achieves a storage saving of 89.2\% (which corresponds to a
deduplication ratio of 9.3$\times$).  

Figure~\ref{fig:storage_vm} shows the storage saving for the VM dataset.
Overall, the storage saving for the first backup reaches 97.4\%
(which corresponds to a deduplication ratio of 38.5$\times$),
mainly because the VM images are initially installed with the same operating
system.  The storage saving drops after the 7th backup, since the students
make big changes and add unique chunks into the VM images. After 13 backups,
the storage saving of the combined scheme achieves 97.9\% (which corresponds
to a deduplication ratio of 47.6$\times$), with a reduction of 0.7\% compared
to that of MLE. 

Overall, the combined scheme maintains high storage efficiency achieved by
deduplication for all datasets.

\subsection{Metadata Access Overhead}
\label{sec:case}

We evaluate the performance of the combined MinHash encryption and scrambling
scheme via a case study of its deployment.  We implement a deduplication
prototype based on the Data Domain File System (DDFS) \cite{zhu08} to simulate
the processing of encrypted deduplication workload.  
DDFS has been used in production backup management for over 15 years
\cite{allu17,douglis17}.  Its chunk locality design also lays the foundation
of various follow-up deduplicated storage systems (e.g.,
\cite{lillibridge09,xia11,ma16}).  Thus, we believe that DDFS is
representative and our study results for DDFS can also be applied to other
locality-based deduplicated storage systems.

Suppose that the chunks have been encrypted, by either the original MLE-based
deterministic encryption or our combined MinHash encryption and scrambling
scheme.  We focus on the metadata access overhead under our DDFS-like
prototype, since metadata access plays an important role in deduplication
performance \cite{zhu08}.

\subsubsection{Prototype Design}

We design and implement our deduplication prototype based on DDFS.
Specifically, our prototype organizes the unique (ciphertext) chunks 
on disk in units of {\em containers} \cite{lillibridge13}.  Each container
size is typically of several megabytes (e.g., 4\,MB) to mitigate the disk seek
overhead, as opposed to the chunk size that is often of several kilobytes
(e.g., 4\,KB or 8\,KB).  In addition, our prototype maintains a 
{\em fingerprint index} to hold the metadata (e.g., the mappings of
fingerprints to chunk locations) and detect if any identical chunk has been
stored.  Since the size of the fingerprint index increases with the amount of
unique chunks being stored, the fingerprint index is stored on disk, while our
prototype maintains two in-memory data structures, namely a {\em fingerprint
cache} and a {\em Bloom filter}, to mitigate the disk I/O overhead during
deduplication (see below). 

Our prototype follows the deduplication workflow of DDFS \cite{zhu08}.  In
particular, it stores unique chunks in logical order and further exploits
chunk locality to accelerate deduplication.  Given an incoming ciphertext
chunk $C$, our prototype performs deduplication as follows. 
\begin{itemize}[leftmargin=*]
\item 
Step~S1: Our prototype checks by fingerprint if $C$ is in the fingerprint
cache. If so, it is identical and does not need to be stored. 
\item 
Step~S2: If $C$ is not in the fingerprint cache, our prototype checks the
Bloom filter.  If $C$ is not in the Bloom filter, it must be unique. Then our
prototype updates the Bloom filter, and also inserts $C$ and its fingerprint
into an in-memory fixed-size buffer in logical order.  If the in-memory buffer
is full, our prototype flushes it to disk as a new container and updates the
fingerprint index on disk. 
\item 
Step~S3: Even if $C$ is in the Bloom filter, it may be a false positive.  Our
prototype queries the fingerprint index to ensure that it is a duplicate.  If
$C$ is not in the fingerprint index, our prototype follows Step~S2 to store
$C$ as a unique chunk.
\item 
Step~S4: If $C$ is in the fingerprint index, our prototype identifies the
container that keeps the physical copy of $C$, and \emph{loads the
fingerprints of all chunks in the container} into the fingerprint cache. The
rationale is that the logically nearby chunks of $C$ are likely to be accessed
together due to chunk locality.  If the fingerprint cache is full, our
prototype removes the least-recently-used fingerprints.
\end{itemize}

\begin{figure}[!t]
\centering
\includegraphics[width=3.3in]{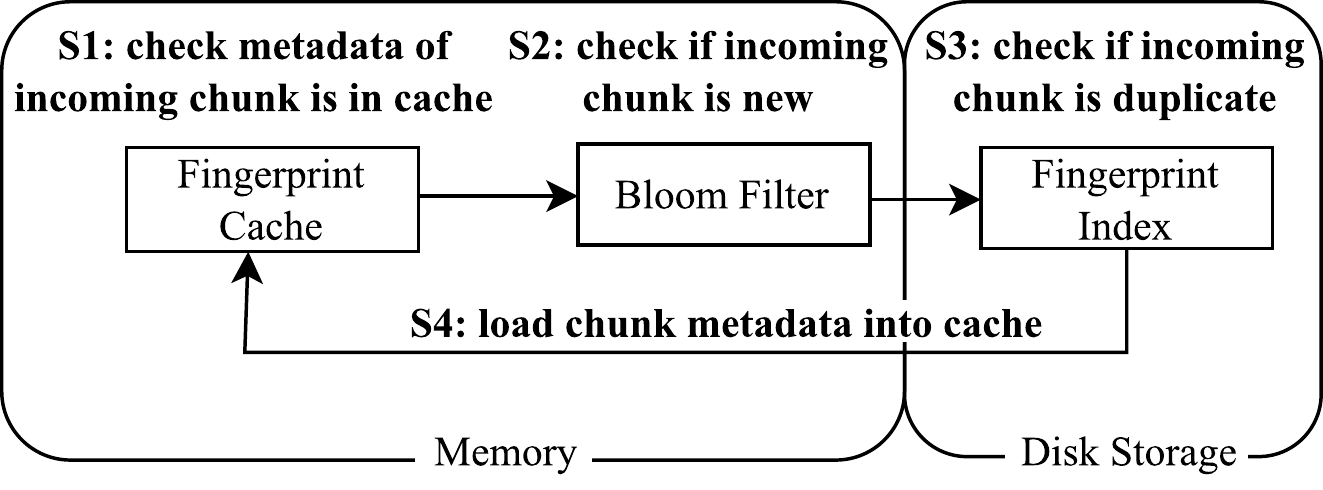}
\caption{Metadata flow of our deduplication prototype.}
\label{fig:case}
\end{figure}

Our prototype mainly implements the metadata flow during deduplication, as
shown in Figure~\ref{fig:case}.  We focus on the evaluation of the metadata
access overhead.  We do not evaluate the performance of writing or reading
containers and that of encrypting or decrypting chunks.  

\subsubsection{Evaluation Results}

Our evaluation uses the following configurations.  Here, we only focus
on the FSL dataset.  We set the metadata size of each fingerprint as 32~bytes.
We consider two sizes of the fingerprint cache: 512\,MB and 4\,GB.  We set the
Bloom filter with a false positive rate of 0.01 \cite{zhu08}, and the Bloom
filter size depends on the number of fingerprints that are tracked.  For
example, our FSL dataset contains around 65~million fingerprints (i.e., the
total size is around 2\,GB), so we use 7 hash functions and the corresponding Bloom filter size is
around 74\,MB.  We also set the container size as 4\,MB. For the combined scheme, we configure its
minimum, average and maximum segment sizes as 512\,KB, 1\,MB and 2\,MB, respectively.
 
We categorize the on-disk metadata access into three types: 
(i) \emph{update access}, which updates the metadata of unique chunks in the
fingerprint index (in Steps~S2 and S3); 
(ii) {\em index access}, which looks up the on-disk fingerprint index for the
detection of duplicate chunks (in Step~S3); and
(iii) \emph{loading access}, which loads the fingerprints of
stored chunks into the cache (in Step~S4).  We measure the metadata access
overhead in terms of the size of metadata being accessed. 

In the following, we compare the metadata access overhead of our combined
MinHash encryption and scrambling scheme with MLE, in which we encrypt the
chunks by the original MLE-based deterministic encryption.   

\begin{figure*}[!t]
\centering
\subfigure[Overall metadata access overhead]{
\includegraphics[height=1.45in]{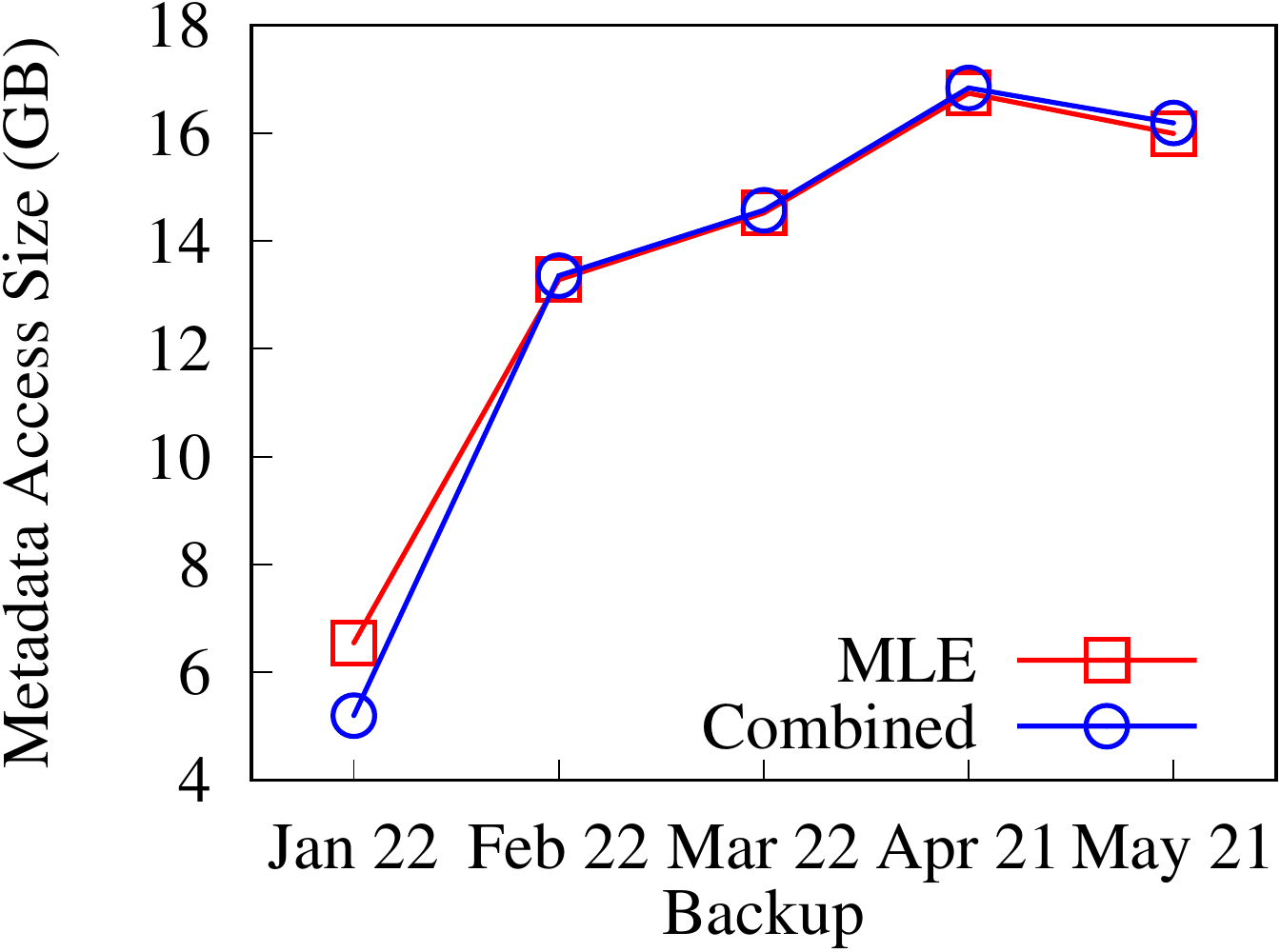}
\label{fig:metadata_fsl_512MB}
}
\subfigure[Breakdown for MLE]{ 
\includegraphics[height=1.45in]{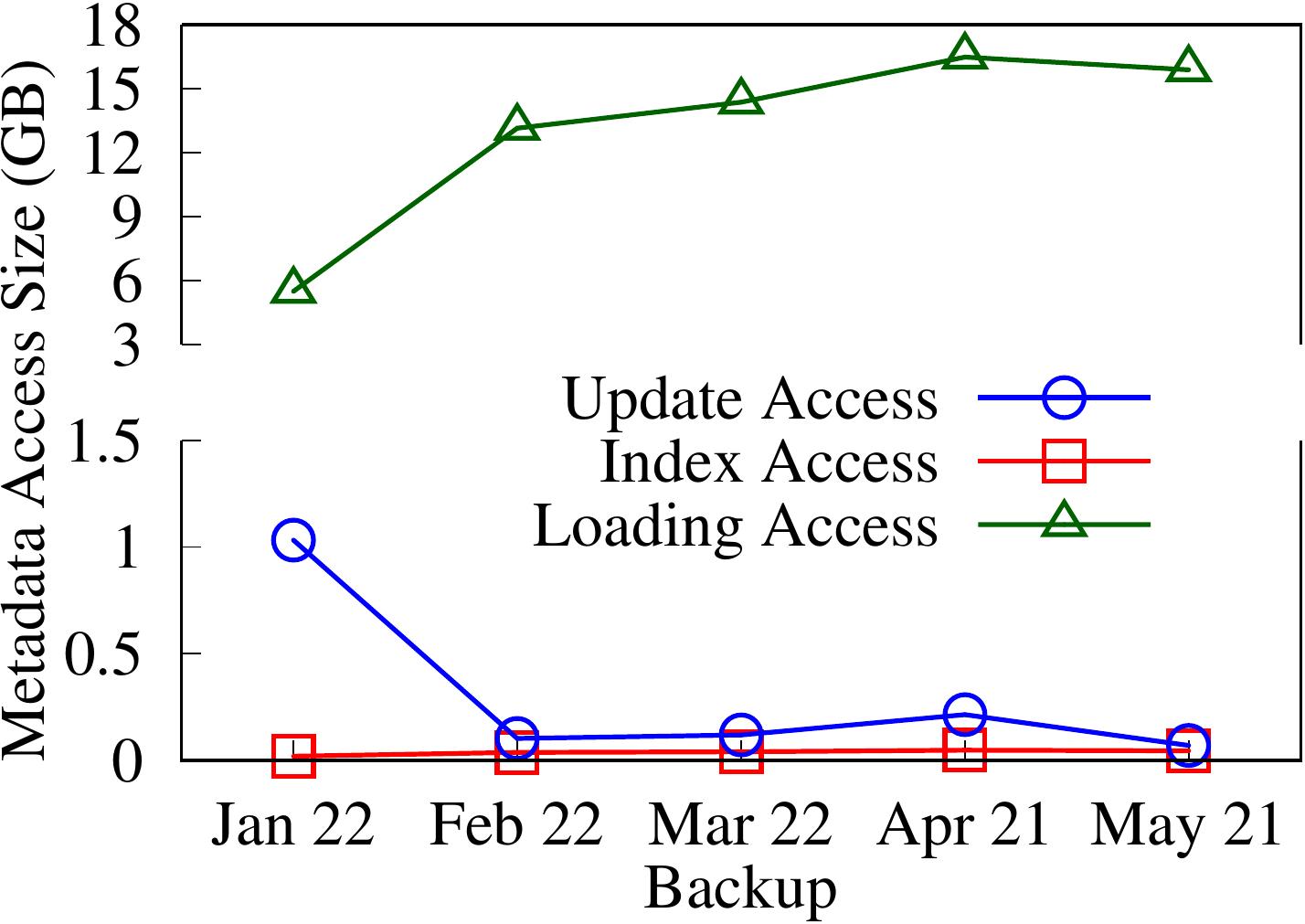}
\label{fig:break_origin_512MB}
}
\subfigure[Breakdown for combined]{
\includegraphics[height=1.45in]{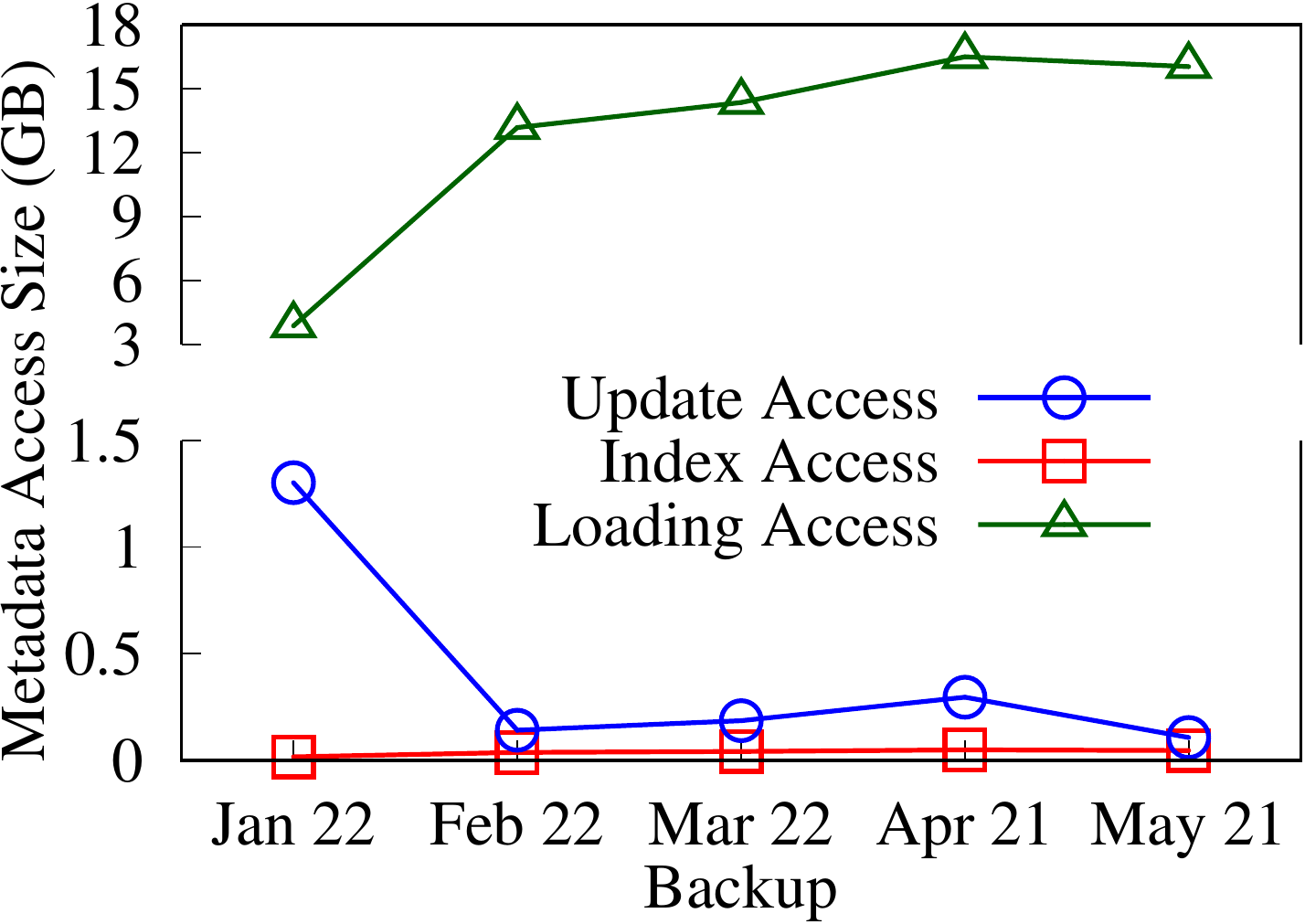}
\label{fig:break_defense_512MB}
}
\vspace{-6pt}
\caption{Performance evaluation: Metadata access overhead when the fingerprint
cache size is 512\,MB.}
\label{fig:meta1}
\end{figure*}

Figure~\ref{fig:meta1} first presents the results when the fingerprint cache
size is 512\,MB, in which case the size is insufficient to hold all fingerprints
in the FSL dataset (whose total metadata size for all fingerprints is around
2\,GB).  Figure~\ref{fig:metadata_fsl_512MB} shows the overall metadata access
overhead.  In the first backup, the combined scheme even incurs less metadata
access overhead than MLE, mainly because it generates more unique chunks at
the beginning and reduces the frequency of loading fingerprints from disk into
the fingerprint cache (in Step~S4).  In the subsequent backups, the combined
scheme has slightly higher overhead than MLE (at most 1.2\%), since it
generates more unique chunks and needs to load fingerprints more often from
disk to the fingerprint cache.  Figures~\ref{fig:break_origin_512MB} and
\ref{fig:break_defense_512MB} show the breakdown of the metadata access
overhead for MLE and the combined scheme, respectively.  The update access
size for both schemes is less than 0.3\,GB after the first backup (in which
MLE and the combined scheme incur 1.0\,GB and 1.3\,GB of metadata access,
respectively), as only a small portion of new or modified chunks are stored.
The index access size is also small, with less than 0.1\,GB for both schemes
in all backups, since a significant portion of duplicate and unique chunks can
be detected by the fingerprint cache and the Bloom filter, respectively. 
Finally, we observe that the loading access size contributes the most
overhead, with more than 74.2\% of the total metadata access size for both
schemes. 
  
\begin{figure*}[!t]
\centering
\subfigure[Overall metadata access overhead]{
\includegraphics[height=1.45in]{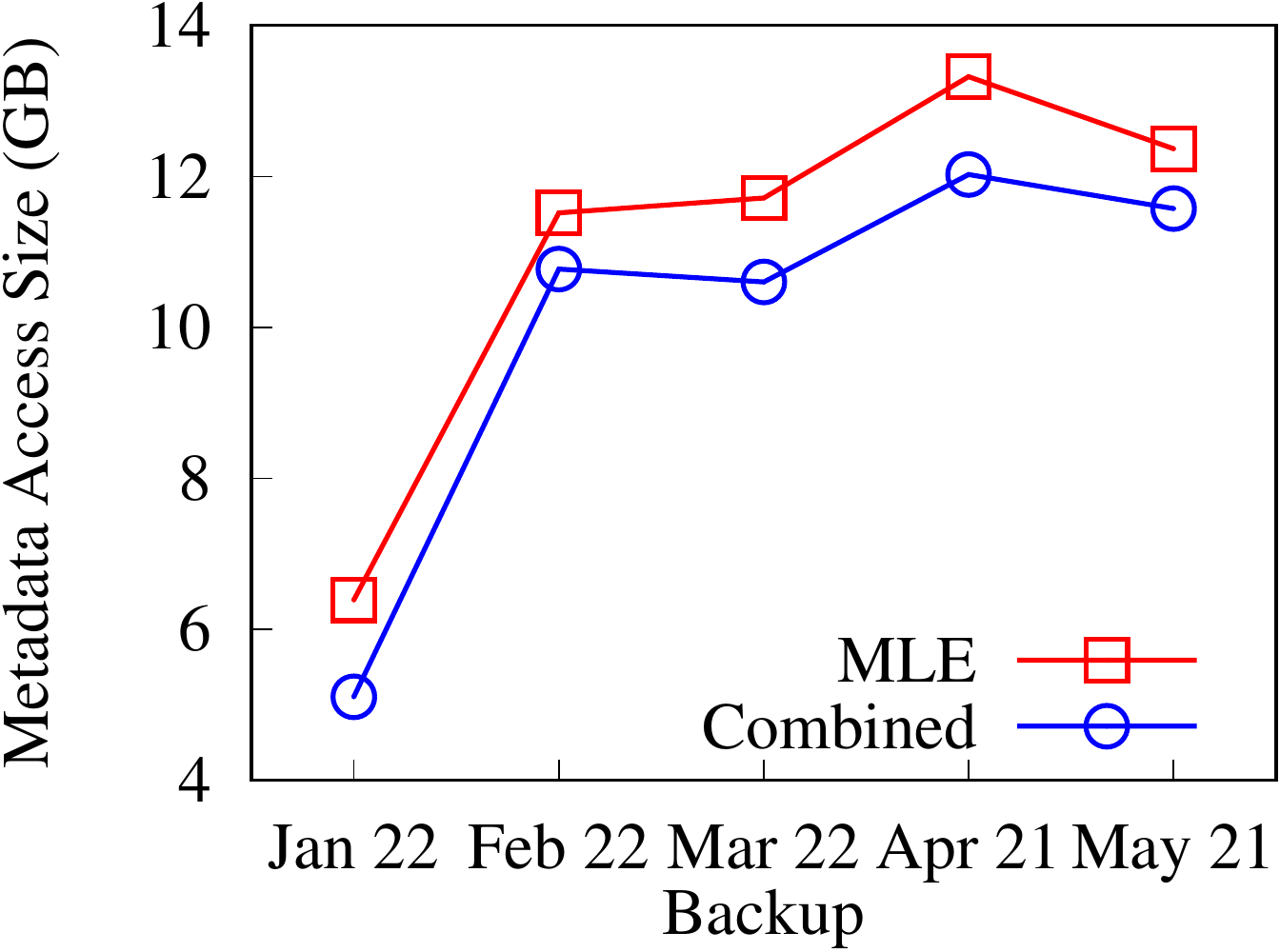}
\label{fig:metadata_fsl_4GB}
}
\subfigure[Breakdown for MLE]{ 
\includegraphics[height=1.45in]{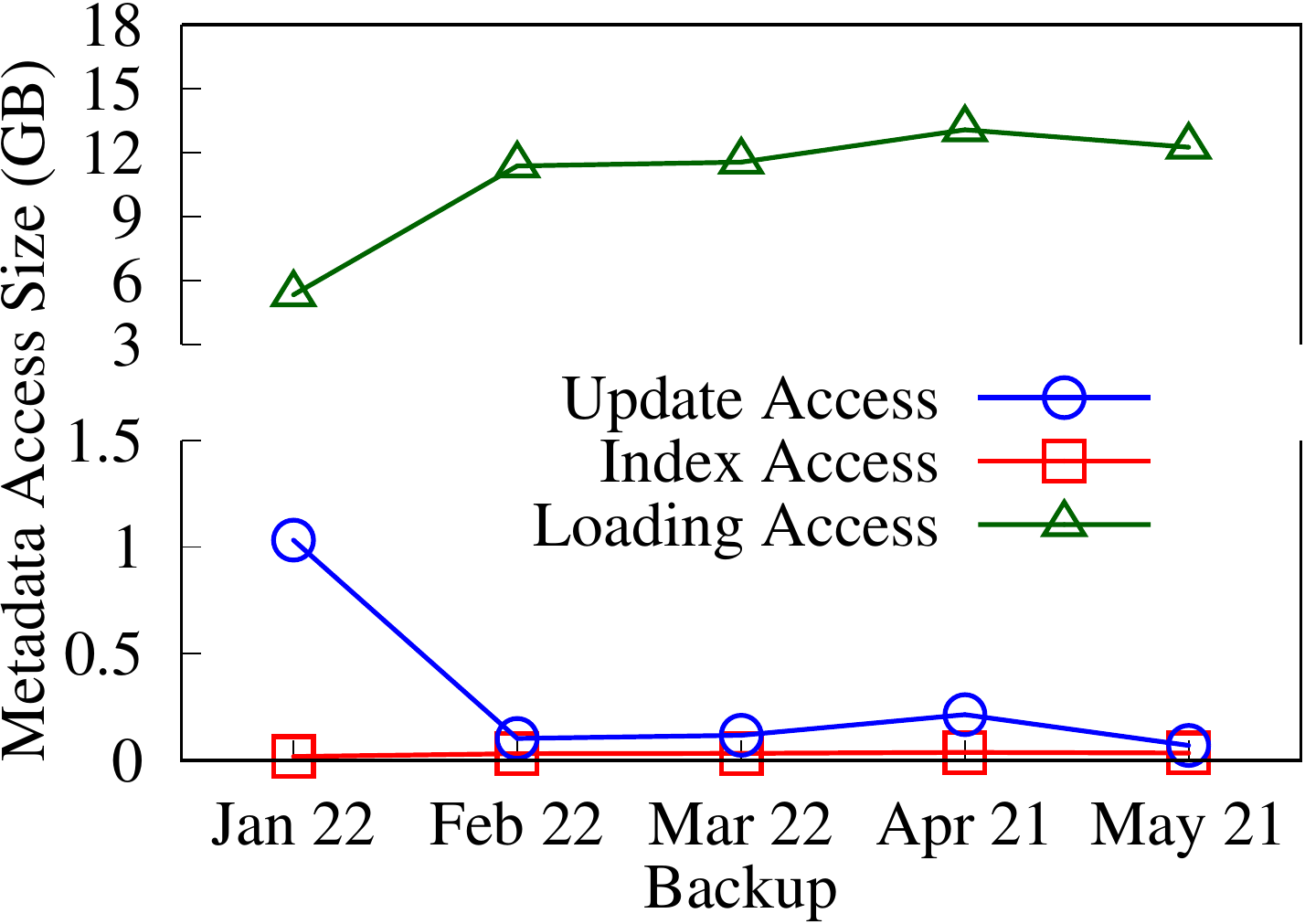}
\label{fig:break_origin_4GB}
}
\subfigure[Breakdown for combined]{
\includegraphics[height=1.45in]{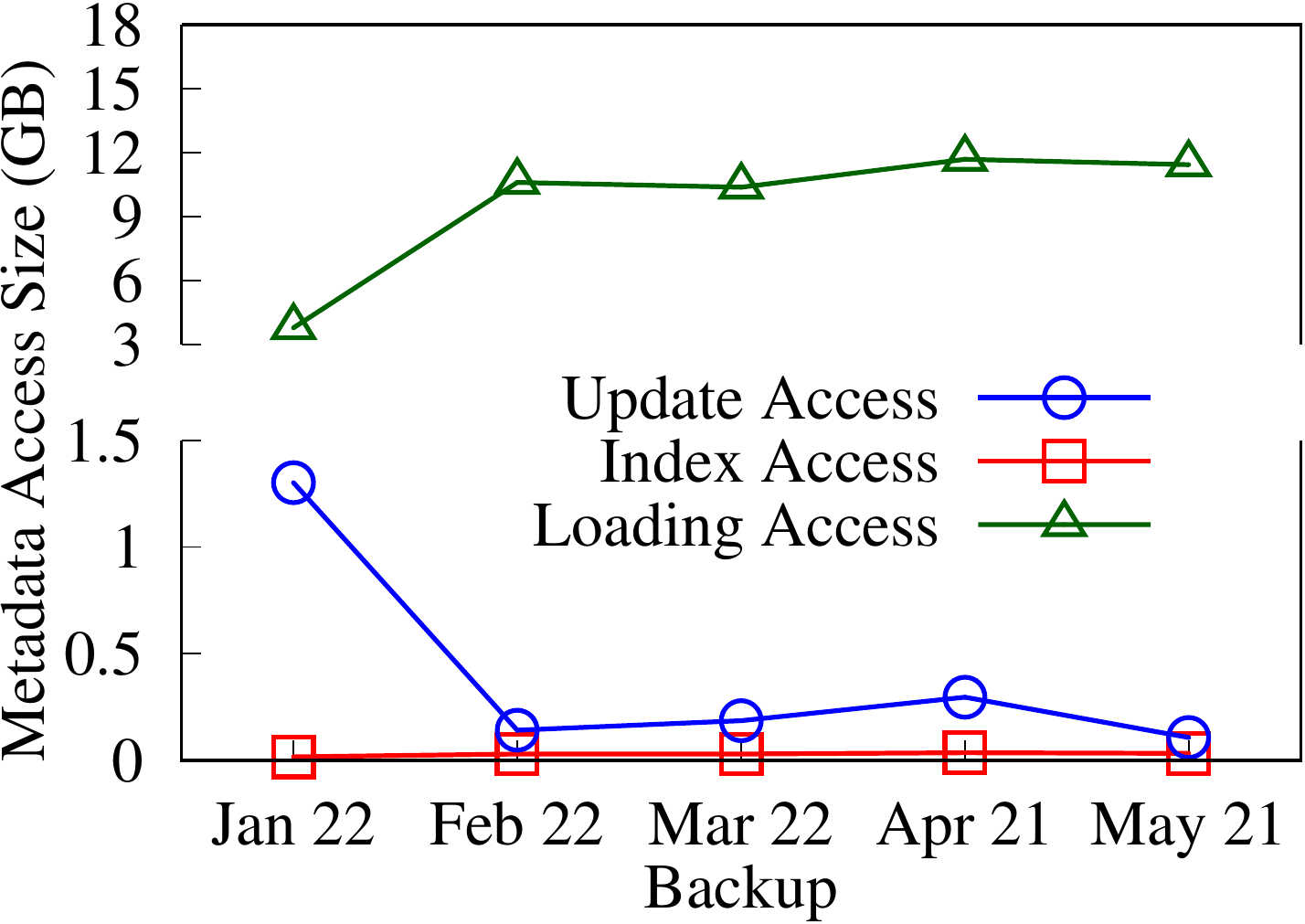}
\label{fig:break_defense_4GB}
}
\vspace{-6pt}
\caption{Performance evaluation: Metadata access overhead when the fingerprint
cache size is 4\,GB.}
\label{fig:meta2}
\end{figure*}

Figure~\ref{fig:meta2} presents the results when the fingerprint cache size is
increased to 4\,GB, in which the fingerprint cache is sufficiently large to
hold the fingerprints of all unique chunks. As shown in
Figure~\ref{fig:metadata_fsl_4GB}, the combined scheme incurs much less
metadata access overhead than MLE by 6.4-20.0\% as it generates more unique
chunks while all fingerprints can be stored in the fingerprint cache. 
Figures~\ref{fig:break_origin_4GB} and \ref{fig:break_defense_4GB} show the
corresponding breakdown for MLE and the combined scheme, respectively.  Both
update access size and index access size are similar to those in
Figure~\ref{fig:meta1}, while the loading access size for both schemes is
significantly reduced by around 22\% and 29\% for MLE and the combined
schemes, respectively, mainly due to a high probability of cache hits. 

\section{Related Work}
\label{sec:related}

\paragraph{Optimizing deduplication:} Existing deduplication
studies (see a complete survey \cite{xia16} on deduplication)
exploit workload characteristics (e.g., chunk locality
\cite{zhu08,lillibridge09,kruus10,xia11} and file similarity
\cite{bhagwat09,xia11}) to mitigate indexing overhead.  For example, DDFS
\cite{zhu08} prefetches the fingerprints of nearby chunks that are likely to
be accessed together.  Sparse Indexing \cite{lillibridge09} and Extreme
Binning \cite{bhagwat09} exploit chunk locality and file similarity,
respectively, to mitigate the memory storage for indexing, while SiLo
\cite{xia11} combines both chunk locality and file similarity for general
backup workloads.  Bimodel \cite{kruus10} builds on chunk locality and
adaptively varies the expected chunk sizes to mitigate metadata overhead. 
All the above works do not consider security.

\paragraph{Encrypted deduplication:} Traditional encrypted deduplication
systems (e.g.,
\cite{douceur02,anderson10,storer08,wilcox-ohearn08,cox02,kallahall02}) mainly
build on convergent encryption \cite{douceur02}, in which the encryption key
is directly derived from the cryptographic hash of the content to be encrypted.
CDStore \cite{li15} integrates convergent encryption with secret sharing to
support fault-tolerant storage. Metadedup \cite{li19} extends CDStore with
space-efficient metadata management.  However, convergent encryption is
vulnerable to brute-force attacks (see Section~\ref{subsec:encrypted-dedup}).
Server-aided MLE protects against brute-force attacks by maintaining
content-to-key mappings in a dedicated key manager, and has been implemented
in various storage system prototypes
\cite{shah15,armknecht15,qin17,bellare13b}.  Given that the dedicated key
manager is a single-point-of-failure, Duan \cite{duan14} proposes to maintain
a quorum of key managers via threshold signature for fault-tolerant key
management.   Note that all the above systems build on deterministic
encryption to preserve the deduplication capability of ciphertext chunks, and
hence are vulnerable to the inference attacks studied in this paper.  

Instead of using deterministic encryption, Bellare {\em et al.}
\cite{bellare13a} propose an MLE variant called {\em random convergent
encryption (RCE)}, which uses random keys for chunk encryption.  However, RCE
needs to add deterministic tags into ciphertext chunks for checking any
duplicates, so that the adversary can count the deterministic tags to obtain
the frequency distribution.  Liu {\em et al.} \cite{liu15}
propose to encrypt each plaintext chunk with a random key, while the key is
shared among users via password-based key exchange.  However, the proposed
approach incurs significant key exchange overhead, especially when the number
of chunks is huge. 

From the theoretic perspective, several studies propose to 
enhance the security of encrypted deduplication and protect the frequency
distribution of original chunks. Abadi {\em et al.} \cite{abadi13} propose two
encrypted deduplication schemes for the chunks that depend on public
parameters, yet either of them builds on computationally expensive
non-interactive zero knowledge (NIZK) proofs or produces deterministic
ciphertext components.  Interactive MLE \cite{bellare15} addresses chunk
correlation and parameter dependence, yet it is impractical for the use of
fully homomorphic encryption (FHE). This paper differs from the above works by
using lightweight primitives for practical encrypted deduplication. 
  
\paragraph{Inference attacks:} Frequency analysis \cite{menezes01} is the
classical inference attack and has been historically used to recover
plaintexts from substitution-based ciphertexts.  It is also used as a building
block in recently proposed attacks. Kumar {\em et al.} \cite{kumar07} use
frequency-based analysis to de-anonymize query logs.  Islam {\em et al.}
\cite{islam12} compromise keyword privacy based on the leakage of the access
patterns in keyword search.  Naveed {\em et al.} \cite{naveed15} propose to
conduct frequency analysis via combinatorial optimization and present attacks
against CryptDB.  Kellaris {\em et al.} \cite{kellaris16} propose
reconstruction attacks against any system that leaks access pattern or
communication volume. Pouliot {\em et al.} \cite{pouliot16} present the graph
matching attacks on searchable encryption.  Grubbs {\em et al.}
\cite{grubbs17} build attacks on order-preserving encryption based on the
frequency and ordering information.

In encrypted deduplication, Ritzdorf {\em et al.} \cite{ritzdorf16} exploit
the size information of deduplicated content and build an inference attack
that determines if a file has been stored. Armknecht {\em et al.}
\cite{armknecht17} present formal analysis on the side-channel attack that
just works in client-side deduplication. Our work is different as we focus on
inferring the content of data chunks via frequency analysis.  In particular, 
we exploit workload characteristics to construct attack and defense
approaches.  

Some inference attacks exploit the \emph{active} adversarial capability.
Brekne {\em et al.} \cite{brekne06} construct bogus packets to de-anonymize IP
addresses. Cash {\em et al.} \cite{cash15} and Zhang {\em et al.}
\cite{zhang16b} propose file-injection attacks against searchable encryption.
Our proposed attacks do not rely on the active adversarial capability.

\section{Conclusion}
\label{sec:conclusion}

Encrypted deduplication has been deployed in commercial cloud environments and
extensively studied in the literature to simultaneously achieve both data
confidentiality and storage efficiency, yet we argue that its data
confidentiality remains not fully guaranteed.  We demonstrate how the
deterministic nature of encrypted deduplication makes it susceptible to
information leakage caused by frequency analysis.  We propose the
locality-based attack, which exploits the chunk locality property of backup
workloads to infer the content of a large fraction of plaintext chunks from
the ciphertext chunks of the latest backup. 
We also propose the advanced locality-based attack, which extends the
locality-based attack with the knowledge of chunk sizes to launch frequency
analysis specifically against variable-size chunks.  We show how the inference 
attacks can be practically implemented, and demonstrate their severities
through trace-driven evaluation on both real-world and synthetic datasets.
To defend against information leakage, we consider MinHash encryption and
scrambling to disturb frequency rank and break chunk locality.  Our
trace-driven evaluation shows that our combined MinHash encryption and
scrambling effectively defends against the locality-based attack, while
maintaining high storage efficiency and incurring limited metadata access
overhead.

\bibliographystyle{plain}
\bibliography{references}

\end{document}